\documentclass[%
superscriptaddress,
secnumarabic,
bibnotes,
 amsmath,amssymb,
 aps,
]{revtex4-2}

\usepackage{graphicx}
\usepackage{dcolumn}
\usepackage{bm}
\usepackage{hyperref}
\usepackage{natbib}
\usepackage[mathlines]{lineno}
\usepackage{xcolor}
\usepackage{tabularx}
\usepackage{booktabs}
\usepackage{threeparttable}

\renewcommand{\thesection}{\Roman{section}}        
\renewcommand{\thesubsection}{\Alph{subsection}}   
\renewcommand{\thesubsubsection}{\arabic{subsubsection}}

\makeatletter
\newcommand{\startappendices}{%
    \appendix
    \setcounter{section}{0}
    \setcounter{subsection}{0}
    \setcounter{subsubsection}{0}
    \setcounter{equation}{0}

    \renewcommand{\thesection}{\Alph{section}}
    \renewcommand{\thesubsection}{\arabic{subsection}}
    \renewcommand{\thesubsubsection}{\arabic{subsubsection}}

    \@addtoreset{equation}{section}

    \renewcommand{\theequation}{\thesection\arabic{equation}}
}
\makeatother

\begin{document}

\preprint{APS/123-QED}

\title{\textbf{Understanding the Theory--Experiment Discrepancy in Pressure Drop of Dilute Polymer Solutions in Channel Flows} 
}%

\author{Nan Hu}
\email{Contact author: nh0529@princeton.edu}
\affiliation{
Department of Mechanical and Aerospace Engineering, Princeton University, Princeton, New Jersey 08544, USA
}

\author{Jonghyun Hwang}
\affiliation{
Department of Mechanical and Aerospace Engineering, Princeton University, Princeton, New Jersey 08544, USA
}

\author{Evgeniy Boyko}
\affiliation{
Faculty of Mechanical Engineering, Technion -- Israel Institute of Technology, Haifa 3200003, Israel
}

\author{Howard A. Stone}
\email{Contact author: hastone@princeton.edu}
\affiliation{
Department of Mechanical and Aerospace Engineering, Princeton University, Princeton, New Jersey 08544, USA
}


\date{\today}

\begin{abstract}
For decades researchers have experimentally observed that the flow of dilute viscoelastic polymer solutions through contraction or contraction--expansion channels yields results at odds with theory and simulations. In particular, the experimentally reported pressure drops are larger than those of generalized Newtonian reference fluids with the same shear viscosity, while constitutive models, such as  Oldroyd-B and FENE-P, predict smaller pressure drops under conditions of low Reynolds numbers and stable flow at small Weissenberg ($Wi$) or Deborah ($De$) numbers. This apparent contradiction between experiments and theory has been a long-standing puzzle in the field. 
Here, we characterize the properties of dilute viscoelastic polymer solutions and employ two distinct types of pressure-sensing systems, conventional recessed pressure taps and flush-mounted diaphragm sensors, to systematically measure pressure drops across channels of different geometrical configurations. These measurements yield qualitative agreement with theoretical predictions across all geometries if the largest relaxation time is adopted for the analysis of the flow. Our results indicate that the apparent discrepancies mentioned above can be attributed to improper interpretation of the measurements and to mismatches between experimental conditions and assumptions made in the theoretical and numerical studies, which include hole pressure effects, the choice of relaxation time of the fluid, and the presence of experimental flow instabilities. 
For quantitative improvements, our results suggest the use of continuum-level constitutive models containing more realistic microscopic features of polymer solutions.

\end{abstract}

\maketitle

\section{Introduction}
Many applications that involve the flow of viscoelastic fluids take place in channels and pipes with varying cross-sectional area, e.g., geometries with contractions, expansions, or contraction--expansion configurations, as occurs in flow in porous media~\cite{Browne2021, Browne2024}, all manners of polymer processing~\cite{Meulenbroek2003}, even viscoelastic electrolytes in batteries~\cite{Janoschka2015,Narayanan2021}, as well as blood flow in stenoses~\cite{Cai2021} and microfluidic devices for handling biofluids~\cite{Sevenler2024}.
The relationship between the pressure drop and the flow rate during such flows is a fundamental problem, which governs both practical performance and physical understanding. However, despite decades of study, this seemingly basic question has remained unresolved due to a long-standing discrepancy between theory and experiments, both for contraction flows \cite{Boger1987,White1987} and for contraction--expansion flows \cite{Rothstein1999,Rothstein2001,Alves2003,CastilloSnchez2022}. Theoretical analyses based on elastic dumbbell constitutive models predict that in contraction or contraction--expansion geometries, low-Reynolds-number flows of viscoelastic polymer solutions should require a smaller pressure drop compared with inelastic fluids of the same shear viscosity~\cite{Szabo1997,Aguayo2008,Keiller1993,Omowunmi2010,Nystrm2012,Zografos2022,Boyko2022,Housiadas2023,Hinch2024,Boyko2024,BoykoStone2024Perspective,Mahapatra2025,kedem2026viscoelastic}.  In contrast, numerous experimental works have reported the opposite trend, namely that dilute polymeric solutions composed of linear polymer chains exhibit an enhanced pressure drop~\cite{Binding1988,Boger1990,Rothstein1999,Rothstein2001,Nigen2002,Miller2009,Ober2013,James2023,Gulati2008,Sousa2011_2,Lanzaro2011}. This contradiction is widely observed across a range of polymeric viscoelastic fluids, such as polystyrene, polyethylene oxide, and polyisobutylene solutions, as well as across different channel geometries, raising a central challenge in rheology and fluid mechanics. Therefore, to bridge this gap, many theoretical studies have attempted to develop new, more complex models, for example, by introducing new dissipative terms \cite{LpezAguilar2016,TamaddonJahromi2016} or accounting for more detailed microscopic dynamics of the dumbbell~\cite{KOPPOL2009,BoykoStone2024Perspective}.  In this work, we explain and understand this discrepancy by providing a new interpretation and understanding from the perspective of experimental measurements, which we further support with our simple representative simulations.

Much of the prior experimental literature has focused on abrupt contraction and contraction--expansion channels, where attention has primarily been directed to vortex growth upstream and downstream of the contraction, associated recirculation zones, and their influence on the flow \cite{Boger1987,McKinley1991,Sousa2011}. These studies commonly attributed the experimentally observed enhanced pressure drop to the presence of vortices. However, numerical simulations have shown that, even when vortices form, the pressure drop should decrease rather than increase \cite{Szabo1997,Alves2003,Omowunmi2010}. This discrepancy indicates that factors beyond the flow structures must be considered, with the pressure drop measurement methodology playing a crucial role.

Several approaches without pressure sensors have been used to determine the “pressure drop.” For instance, in several studies, viscoelastic fluids have been driven using a piston \cite{Nystrm2016} or compressed gas \cite{James2021,Nigen2002,James2023}, interpreting the applied pressure as the pressure drop. However, such methods introduce various uncertainties, including piston-wall friction and the presence of a not fully developed flow. Importantly, the channel-exit pressure of a viscoelastic fluid into air cannot simply be assumed equal to the ambient atmospheric pressure \cite{Han1993,Padmanabhan1994}. This highlights the necessity of pressure sensors to measure the pressure drop of viscoelastic fluids when considering the question at hand.

The most common measurement approach involves placing pressure taps upstream and downstream of the test section and connecting them to differential or independent pressure transducers. The resulting pressure difference, either the direct differential reading~\cite{James1980,James1982,Cartalos1992,Rodd2005,Rodd2007,Miller2009,Sousa2009,Sousa2010,Sousa2011_2,Lanzaro2011} or the difference between two gauges~\cite{Li2011,PrezCamacho2015,Gulati2008,Rodd2010}, is then reported as the pressure drop. However, such measurements {for viscoelastic polymer solutions} inevitably introduce the so-called ``hole pressure effect'', which has been studied extensively and ought to be well understood \cite{BROADBENT1968,Higashitani1972,Novotny1973,bird1987dynamics1}, yet it appears to not always be given consideration or verification in this context \cite{James1980,James1982,Cartalos1992,Rodd2005,Rodd2007,Miller2009,Sousa2009,Sousa2010,Li2011,PrezCamacho2015}. To the best of our knowledge, only a few studies~\cite{Binding1988,James1990,James1990_2,Rothstein1999,Rothstein2001,Ober2013,Keshavarz2016} have employed flush-mounted diaphragm sensors. In particular, \citet{Binding1988}~examined planar and axisymmetric contractions and reported an increased pressure drop. However, their setup discharged through an orifice to the atmosphere and used the pressure at the base of the contraction wall as the reference, which was dominated by the vortex region. This reported ``pressure drop'' is therefore not the pressure drop defined in theoretical or numerical studies. For axisymmetric hyperbolic contractions, \citet{James1990,James1990_2} reported a decreased pressure drop compared with an inelastic fluid, in qualitative agreement with theoretical predictions. However, this finding has not received sufficient attention.

Two experimental studies~\cite{Ober2013,Keshavarz2016} with 
Boger fluids and 
using flush-mounted microsensors reported increased pressure drop, but the Reynolds number was $Re=\mathcal{O}(1$–-$10^2)$. Also, for $Re\ll1$,~\citet{Rothstein1999,Rothstein2001} conducted careful experiments on axisymmetric contraction--expansion flows and reported an increased pressure drop prior to the onset of instability. In their analysis, however, two different relaxation times $\lambda$ were reported: a larger value $\lambda_{z}$ extracted from Zimm-model fitting of the storage modulus $G'$ and loss modulus $G''$ from small-amplitude-oscillatory-shear (SAOS) measurements, and a much smaller value $\lambda_{\Psi1,0}$ obtained from calculation of the zero-shear rate first normal stress coefficient $\Psi_{1,0}=2 \lim _{\omega \rightarrow 0} \frac{G^{\prime}(\omega)}{\omega^2}$~\cite{Rothstein1999}. They adopted $\lambda_{\Psi1,0}$ to define the Weissenberg number ($Wi_{\Psi1,0}$), thereby placing their conclusions in a nominally low-$Wi$ regime, namely $Wi_{\Psi1,0}\lesssim2.5$ corresponding to a stable regime, where the flow would be characterized with $Wi_{z}\lesssim55$ if $\lambda_z$ is adopted.
However, the appropriate choice of relaxation time has long been debated \cite{Boger1992,James2021}. For example, \citet{James2021} argued that $\lambda_{\text{caber}}$, based on capillary-breakup-extensional rheometry (CaBER), is more representative than those $\lambda_{N1}$ inferred from the first normal stress difference $N_1$,  where $N_1$ is measured from the steady shear flow. 
In our work, we also find that $\lambda_{\text{caber}} \approx \lambda_{z} \gg \lambda_{N1}$, consistent with the literature \cite{Sousa2009,Sousa2011b,James2021,Gaillard2025} (see Appendix~\ref{appendix_relaxation_time} and Table~\ref{table_relaxtion_time}). This large disparity implies that the dimensionless parameter range of $Wi$ should be carefully considered when comparing numerical and theoretical predictions with experimental measurements \cite{KOPPOL2009}.

Therefore, in this work,
we employ two distinct types of pressure-sensing systems—conventional recessed pressure taps and flush-mounted diaphragm sensors—to systematically measure pressure drops across channels of different geometrical configurations, using carefully prepared dilute polyisobutylene (PIB) solutions as representative fluids, which are linear polymer solutions consistent with those reported in the literature (see Table~\ref{table_summary}), ensuring a fair comparison. By directly comparing the responses of these two types of sensors, we identify that in flows involving only contraction or only expansion, the use of a flush-mounted sensor is essential. In contrast, in contraction--expansion flows, placing recessed pressure sensors sufficiently far upstream and downstream of the contraction--expansion region can avoid the hole pressure effect. However, because the resulting measurement includes the much larger pressure drop from the long straight inlet and outlet channels, the local asymmetric viscoelastic pressure response of the contraction--expansion section may be easily masked. Furthermore, we demonstrate that the theoretical predictions and experimental observations are qualitatively consistent in almost all cases, effectively eliminating the previously claimed discrepancies. Nevertheless, we recognize that adopting different relaxation times in the characterization of the fluid can shift the values of non-dimensional Weissenberg and Deborah numbers, which not only complicates quantitative predictions but may also lead to inconsistent interpretations.

\begin{figure*}[t!]
\centering
\includegraphics[width=\linewidth]{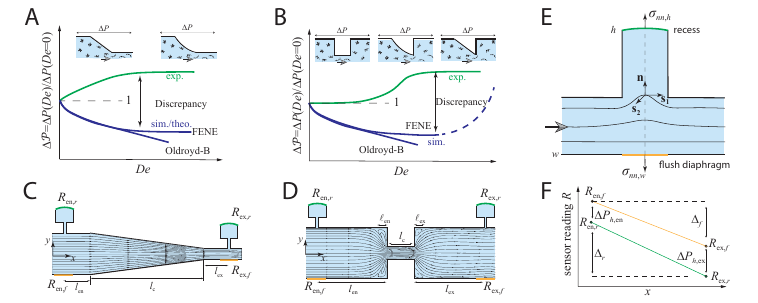}
\caption{Schematic illustration of the commonly observed discrepancy between theoretical and experimental pressure drops in viscoelastic flows through variable cross-section channels, and of the hole pressure effect in pressure measurements. 
(A)–(B) Conceptual comparison of theoretical and experimental results for contraction and contraction--expansion flows, where the dashed lines indicate that $\Delta \mathcal{P}$ may exceed unity at sufficiently high $De$ for both planar and axisymmetric configurations  \cite{Szabo1997,Aguayo2008,KOPPOL2009,Zografos2020}. Our work reported here suggests that trends in (A) may reflect measurement artifacts, while improper measurements or interpretation mismatches may exist between experimental and theoretical parameter ranges in (B).
(C)–(D) Experimental pressure measurements for the same geometries, where the recessed sensor (green curves) represents the conventional method, in which the liquid column within the hole transmits the pressure to the diaphragm, whereas the flush sensor (orange lines) measures the in situ pressure directly. Sensor readings $R$ use subscripts “en”, “ex”, “$r$”, and “$f$” to denote entrance, exit, recessed, and flush configurations, respectively. 
(E) Streamline geometry near recessed and flush sensors, with $\mathbf{s}$, $\mathbf{n}$, and $\mathbf{b}$ indicating the tangential, normal, and binormal directions. The sensors effectively probe different normal stresses $\sigma_{\it nn}$ at the wall surface (w) or inside the hole (h). 
(F) Schematic of sensor readings along the channel. The difference $\Delta$ between upstream and downstream readings is often interpreted as the pressure drop $\Delta P$, but actually corresponds to the normal stress difference $\Delta \sigma_{\it nn}$.}
\label{fig1}
\end{figure*}

\section{Long-standing discrepancy}

A discrepancy between experiments and theory for the pressure drop--flow rate ($\Delta P$–$Q$) behavior of dilute viscoelastic polymer solutions in variable-cross-section channels is illustrated in Fig.~\ref{fig1}A and \ref{fig1}B. Fig.~\ref{fig1}A illustrates a (abrupt or smooth) contraction, and Fig.~\ref{fig1}B shows an abrupt contraction--expansion. In both geometries, 
and in low-Reynolds-number conditions ($Re \ll 1$), measurements reported in the literature consistently show that as the volumetric flow rate $Q$ increases, the steady pressure drop of a viscoelastic fluid exceeds that of an inelastic fluid with the same shear viscosity~\cite{Binding1988,Boger1990,Rothstein1999,Rothstein2001,Nigen2002,Miller2009,Ober2013,James2023,Gulati2008,Sousa2011_2,Lanzaro2011}. This observation is commonly expressed in terms of a normalized pressure drop
\begin{equation}
\Delta\mathcal{P}(Q)\;\equiv\;\frac{\Delta P(De;Q)}{\Delta P(De=0;Q)},
\end{equation}
where $\Delta P(De;Q)$ is the pressure drop of a viscoelastic fluid at Deborah number $De$, and $\Delta P(De=0;Q)$ is the pressure drop for a viscous (non-elastic) fluid with the same shear viscosity. The Deborah number is defined here as
$De \;\equiv\; \lambda Q/(l_c A)$,
with $\lambda$ the fluid relaxation time, $l_c$ a characteristic length of the contraction or throat, and $A$ the minimum  cross-sectional area of the channel. The Deborah number $De$ is related to the Weissenberg number $Wi \equiv \lambda \dot{\gamma}$ through an appropriate geometric prefactor involving the channel aspect ratio~\cite{Boyko2022}, where $\dot{\gamma}$ is a characteristic shear rate. The inequality $\Delta\mathcal{P} > 1$ has been reported experimentally both for nearly constant shear-viscosity, viscoelastic (Boger) polymer solutions \cite{Binding1988,Boger1990,Rothstein1999,Rothstein2001,Nigen2002,Miller2009,Ober2013,James2023} and for strongly shear-thinning, viscoelastic  polymer solutions \cite{Gulati2008,Sousa2011_2,Lanzaro2011}. Further, $\Delta\mathcal{P} > 1$ persists largely independently of the specific channel geometry, including contraction ratio, abruptness, contour of the channel, and parameter range (e.g., $Re,De,Wi$), as summarized in Table~\ref{table_summary}.

In contrast, analyses based on widely employed constitutive equations, such as the Oldroyd-B and FENE (finitely extensible
nonlinear elastic) models, predict the opposite trend to the experimental measurements. Both numerical simulations \cite{Keiller1993,Szabo1997,Aguayo2008,TamaddonJahromi2016,Keiller1993,Omowunmi2010,Nystrm2012,Zografos2020,Zografos2022,Hinch2024,Mahapatra2025} and asymptotic solutions~\cite{Boyko2022,Housiadas2023,Hinch2024,Boyko2024,Mahapatra2025} indicate that, within these contraction--expansion \cite{Szabo1997,Aguayo2008,TamaddonJahromi2016,Zografos2020,kedem2026viscoelastic} or contraction geometries~\cite{Keiller1993,Omowunmi2010,Nystrm2012,Boyko2022,Housiadas2023,Hinch2024,Boyko2024,Mahapatra2025,Zografos2022}, viscoelastic fluids should exhibit $\Delta\mathcal{P} < 1$  when $De$ is small (Table~\ref{table_simulation_summary}), though $\Delta\mathcal{P} > 1$ may occur at large $De$ for contraction--expansion \cite{Szabo1997, Aguayo2008, KOPPOL2009,Zografos2020}. For the contraction geometry, the Oldroyd-B model predicts a monotonic
pressure drop reduction at low and high $De$~\cite{Hinch2024,Boyko2024}, whereas the FENE model \cite{Mahapatra2025}, which accounts for finite polymer extensibility, predicts leveling off of the pressure drop to a plateau at sufficiently large $Q$ (or high $De$), followed by a slight increase in $\Delta\mathcal{P}$. Nevertheless, even at sufficiently high Deborah numbers, the FENE model predicts $\Delta\mathcal{P} < 1$.

As noted by~\citet{Hinch2024} and~\citet{Boyko2024}, the pressure drop reduction predicted by the Oldroyd-B model in contraction flow arises from two effects. First, the elastic normal stresses, representing the tension in the streamlines, are larger at the end of the contraction than at the entrance, effectively pulling the fluid forward and reducing the required pressure. Second, the elastic shear stresses relax only far downstream, so their contribution remains below the steady Poiseuille value, further decreasing the pressure drop.

\section{Pressure measurements in viscoelastic fluid flows and the hole pressure effect}

Classical pressure-drop measurements in pipe or channel flows are typically obtained using small sidewall holes connected to recessed pressure transducers (green curves in Fig.~\ref{fig1}C,D), yielding upstream and downstream readings $R_{\mathrm{en},r}$ and $R_{\mathrm{ex},r}$. In contrast, using a flush diaphragm sensor (orange curves in Fig.~\ref{fig1}C,D), whose surface is aligned with the channel wall, provides the corresponding readings $R_{\mathrm{en},f}$ and $R_{\mathrm{ex},f}$.

For Newtonian liquids, pressure taps formed by drilled side holes are generally reliable, as the perturbation to the local stress field is negligible. For viscoelastic fluids, however, the local kinematics near the hole can significantly disturb the stress field as a consequence of elastic effects \cite{BROADBENT1968,Higashitani1972,Novotny1973}. A schematic comparison of the local stress transmission and streamline geometry for the two sensor types is provided in Fig.~\ref{fig1}E. In fact, the diaphragm membranes of the recessed and flush sensors probe different local normal stresses, $\sigma_{\it nn}$, because of differences in streamline curvature associated with the two distinct configurations, 
$\sigma_{nn,h}$ and $\sigma_{nn,w}$, where the subscripts “$h$" and “$w$" denote the hole and wall, respectively. Based on a local surface coordinate system, the divergence of the stress may be expressed as a balance of projections along two local tangent directions, $\mathbf{s}_1$ and $\mathbf{s}_2$ (unit vectors), and the surface-normal direction $\mathbf{n}=\mathbf{s}_1\times\mathbf{s}_2$. At the centerline of a circular pressure tap, we assume that the tangential variations vanish locally, $\partial(\cdot)/\partial s_1=\partial(\cdot)/\partial s_2=0$, and that the local geometry is axisymmetric, such that the two principal curvatures are equal, $\kappa_1=\kappa_2=\kappa$. The divergence of the stress tensor $\boldsymbol{\sigma}$ in a low-Reynolds-number flow with negligible fluid inertia then reduces to (see Appendix~\ref{appendix_hole_pressure} for details) 
\begin{equation} \label{EqnStressDivergence} \begin{aligned} \boldsymbol{0} =\boldsymbol{\nabla}\cdot\boldsymbol{\sigma} = \left( \frac{\partial\sigma_{1n}}{\partial n} +3\kappa\sigma_{1n} \right)\mathbf{s}_1 + \left( \frac{\partial\sigma_{2n}}{\partial n} +3\kappa\sigma_{2n} \right)\mathbf{s}_2 + \left[ \frac{\partial\sigma_{nn}}{\partial n} +\kappa(\sigma_{nn}-\sigma_{11}) +\kappa(\sigma_{nn}-\sigma_{22}) \right]\mathbf{n}. \end{aligned} \end{equation} This formulation shows that the local stress variations near the sensor depend explicitly on the local surface curvature $\kappa$.
By applying the Oldroyd--B constitutive model to Eq.~\eqref{EqnStressDivergence} and assuming a predominantly unidirectional flow along $\mathbf{s}_1$, such that $\sigma_{2n}\simeq0$, while eliminating the curvature through the tangential and normal momentum balances (see Appendix~\ref{appendix_hole_pressure} for details), the difference between the flush and recessed sensor measurements, $\Delta P_h$, reduces to \begin{equation} \Delta P_h \equiv P_w-P_h = \sigma_{nn,h}-\sigma_{nn,w} = \frac{1}{6}N_{1,w}, \label{3d_results} \end{equation} where $P=-\sigma_{nn}$ and $N_{1,w}$ is the first normal-stress difference at the wall. 
In particular, for an Oldroyd--B fluid in pressure-driven flow, $ N_{1,w} = 2\eta_p\lambda\dot{\gamma}_w^2 \propto Q^2 $, where $\eta_p$ is the polymeric viscosity, $\lambda$ is the relaxation time, and $\dot{\gamma}_w$ is the wall shear rate. Eq.~\eqref{3d_results}, first reported in a general tensorial formulation using Christoffel symbols~\citep{Higashitani1972}, shows that the difference measured between a pair of ``flush--recessed'' sensors increases quadratically with the shear rate, or equivalently with the flow rate $Q$. Here, we obtain the same result using a more geometrically transparent formulation based on local surface coordinates. This provides a concise theoretical explanation for the hole-pressure effect expected in viscoelastic pressure measurements.

When a viscoelastic fluid flows through a channel of varying cross section, the local shear rates upstream and downstream of the sensors differ, leading to substantial differences between recessed and flush readings, denoted $\Delta_r$ and $\Delta_f$. \emph{This distinction is crucial because previous studies have often interpreted $\Delta_r$ as the pressure drop \cite{Sousa2009,Sousa2011b}, although it also reflects upstream--downstream elastic stress differences.} 
As illustrated in Fig.~\ref{fig1}F for a contraction flow, the downstream sensor experiences a higher shear rate, yielding a larger hole-induced offset ($\Delta P_{h,\mathrm{ex}} > \Delta P_{h,\mathrm{en}}$). Consequently, the recessed measurement $\Delta_r$ overestimates the true pressure drop, which should instead be taken as the flush-sensor value $\Delta_f$, i.e., $\Delta P = \Delta_f < \Delta_r$.

\begin{figure}
\centering
\includegraphics[width=0.6\linewidth]{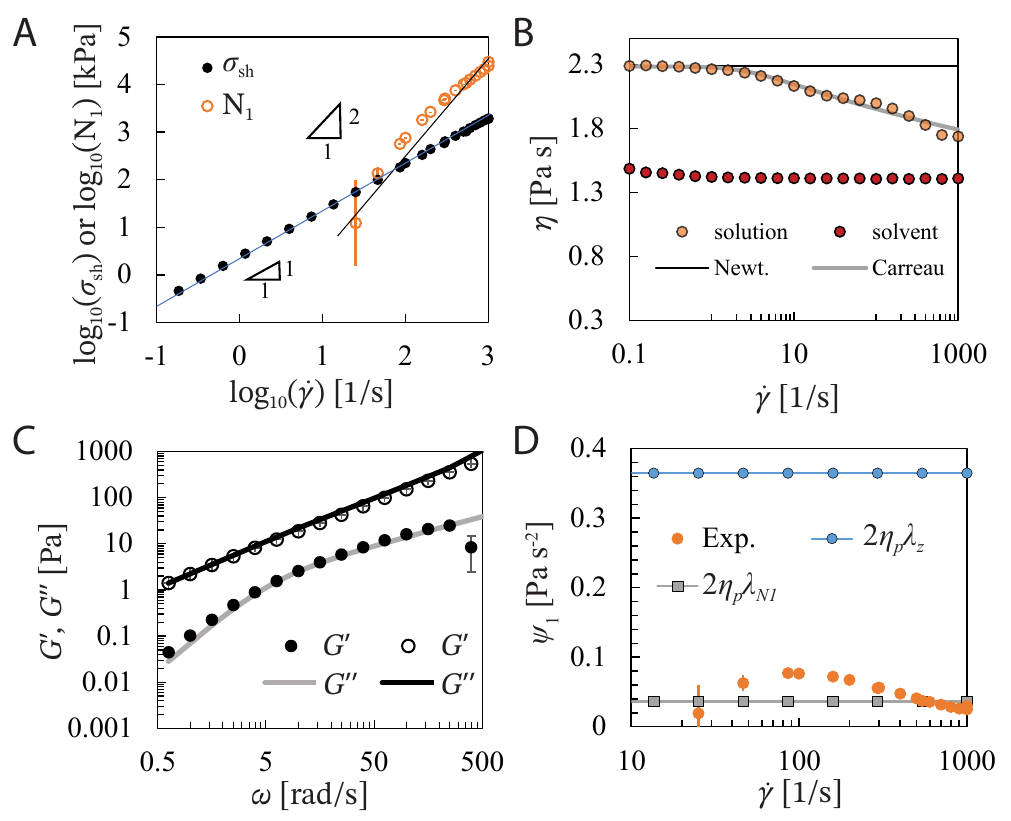}
\caption{Rheological characterization of the test solution, consisting of 0.5~wt.\% PIB dissolved in a mixture of PB and mineral oil. (A) Measured shear stress $\sigma_{\mathrm{sh}}$ and first normal-stress difference $N_1$ as a function of shear rate $\dot{\gamma}$.
(B) Shear viscosity $\eta$ as a function of shear rate $\dot{\gamma}$. The gray curve shows the fit with the Carreau model and the black line indicates the Newtonian reference. 
(C) Storage modulus ($G'$) and loss modulus ($G''$) as a function of oscillatory frequency $\omega$ at 1\% strain. Gray and black curves denote fits with the Zimm model. (D) Measured first normal-stress coefficient $\Psi_1 = N_1/\dot{\gamma}^2$ and the predicted value $2\eta_p\lambda$ based on $\lambda_z$ and $\lambda_{N1}$.
Error bars are shown for all data; where not visible, they are smaller than the symbol size.}
\label{fig2}
\end{figure}

\section{Experimental measurements of various geometries}
\subsection{Experimental measurements and pressure drop in straight channels}

We performed experiments with a polymer solution consisting of polyisobutylene (PIB) dissolved in polybutene (PB) and oil \cite{More2023,Hu2025,2025Jong}. This rheologically stable system allows the viscosity to be tuned to approximate a Boger fluid~\cite{james2009boger}. Several PIB concentrations were tested (Appendix~\ref{appendix_fluids}, Fig.~\ref{figs_rheo_PIB_PB}), and 0.5~wt.\% PIB was selected because it remains in the dilute regime (Fig.~\ref{figs_rheo_PIB_PB}A) while providing a sufficiently strong viscoelastic response for pressure sensing (Fig.~\ref{figs_rheo_PIB_PB}B).

The rheological properties of the chosen solution are shown in Fig.~\ref{fig2}, including the first normal-stress difference $N_1$, shear stress $\sigma_{\mathrm{sh}}$, viscosity $\eta(\dot{\gamma})$, and the dynamic moduli ($G',G''$). The fluid behaves approximately as a Boger fluid, with $\sigma_{\mathrm{sh}}\!\propto\!\dot{\gamma}$ and $N_1\!\propto\!\dot{\gamma}^2$ over several decades of shear rate (Fig.~\ref{fig2}A), but exhibits slight shear thinning (Fig.~\ref{fig2}B). Even weak shear thinning affects pressure drop measurements, so the viscosity was fit using the Carreau model (gray curve, Fig.~\ref{fig2}C) to define the inelastic reference.
As a dilute $\theta$-solution, the dynamic moduli follow the Zimm model \cite{More2023,2025Jong}, yielding a relaxation time $\lambda_z = 0.227$~s (Appendix~\ref{appendix_relaxation_time}\ref{appendix_zimm}) consistent with $\lambda_{\text{caber}}=0.224$~s from capillary-breakup rheometry (Appendix~\ref{appendix_relaxation_time}\ref{appendix_caber}). A smaller relaxation time, $\lambda_{N1} = \Psi_1/(2\eta_p) = 0.0228$~s, is obtained from the first normal-stress coefficient (Fig.~\ref{fig2}D or Appendix~\ref{appendix_relaxation_time}\ref{appendix_N1}), consistent with the typical ordering $\lambda_z\!\approx\!\lambda_{\text{caber}}\!\gg\!\lambda_{N1}$ for dilute viscoelastic polymer solutions \cite{Rothstein1999,James2021,Gaillard2025}. This experimental observation has been widely found and discussed (Table~\ref{table_relaxtion_time}), as we also documented and discussed in Appendix~\ref{appendix_relaxation_time}\ref{appendix_relaxation_discuss}. Despite it still being debated, both $\lambda_z$ and $\lambda_{N1}$ are therefore used in later simulations and non-dimensional analysis in this work. {It is also noted that the experimentally measured first normal-stress coefficient $\Psi_1$ is not monotonic in shear rate. This behavior is observed not only in the primary fluid used in this work, but also in the additional fluid reported in the Appendix~\ref{appendix_another_fluid}, in our recent systematic study of Boger fluids \cite{2025Jong}, and in the work of James \& Roos \cite{James2021}. The real difficulty is that no simple constitutive model considered here can simultaneously and quantitatively predict all rheological properties of a real dilute polymer solution (also see the discussion of Fig.~\ref{mFENE_CR_features} in Appendix~\ref{appendix_simulation}2), even though such fluids are often regarded as relatively simple model viscoelastic fluids. Nevertheless, we consider that simple dumbbell models remain useful because they can still provide qualitative predictions and help identify the dominant physical mechanisms. }

\begin{figure*}[t!]
\centering
\includegraphics[width=\linewidth]{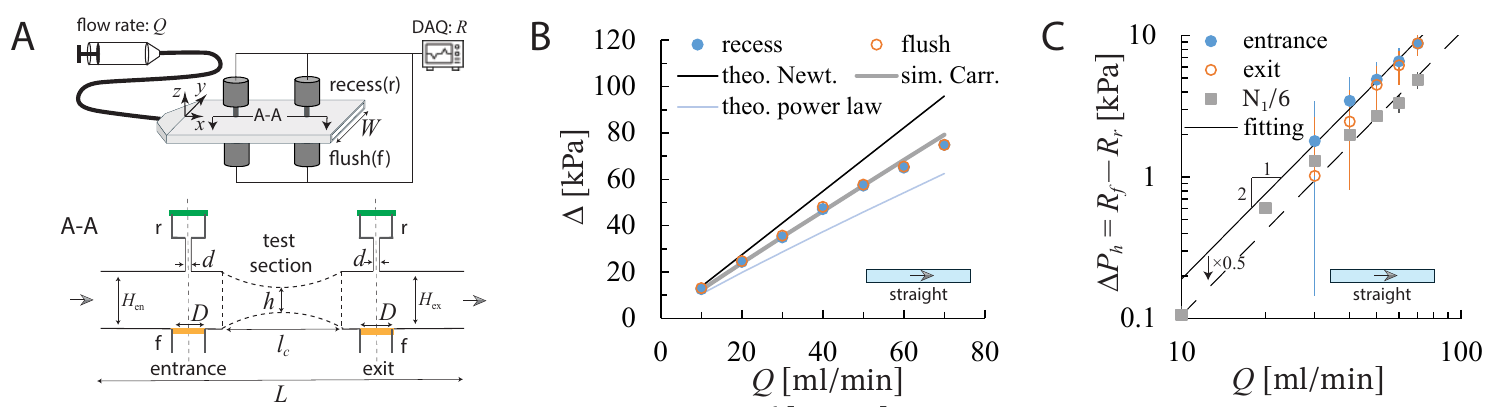}
\caption{Schematic of the pressure-drop measurement approach and results in a straight channel. 
(A) Experimental setup: a syringe pump supplies a constant flow rate $Q$, and flush- and recess-mounted pressure sensors are placed at the upstream and downstream ends of the test section. Sensor readings $R$ are collected using a data acquisition (DAQ) unit. The planar channel satisfies $h \ll W \ll L$. 
(B) Comparison of the sensor reading differences $\Delta$ with theoretical and numerical predictions for a straight channel of geometry \{$H,h,l_c$\}=\{$1,1,30$\}~mm. Orange circles indicate $\Delta_f$ from flush sensors, and blue dots indicate $\Delta_r$ from recessed sensors. The gray curve shows Carreau-model simulations based on the measured shear viscosity, while the black and blue curves correspond to Newtonian ($\eta=\eta_0$) and power-law approximations, respectively. 
(C) Hole pressure effect quantified by $\Delta P_h = R_f - R_r$. Gray markers show Eq.~\eqref{3d_results}, where $N_1$ is obtained from the rheometric measurements. Solid line is fitted from $\Delta P_{h,\text{en}}$. The dashed line is a guideline to indicate the trend.}
\label{fig3}
\end{figure*}

The experimental setup for pressure drop measurements is shown in Fig.~\ref{fig3}A, whose details can be found in Appendix~\ref{appendix_exp}. A syringe pump drives the viscoelastic fluid at a constant flow rate $Q$ through a planar channel of width $W$, length $L$, and height $H$ (or minimum throat height $h$). Recessed and flush sensors were mounted opposite each other at the inlet and outlet, and all four signals were recorded at 20~Hz. The channels were fabricated by 3D printing with a nominal resolution of $25~\mu$m and fixed dimensions $W=10$~mm, $H=4$~mm, $h=1$~mm, and $L=90$~mm, with 30-mm entrance and exit regions to ensure fully developed flow. The maximum shear rate in the throat is $\dot{\gamma}_w = 6Q/(Wh^2)\approx 10^3~\mathrm{s^{-1}}$ at $Q=100~\mathrm{ml/min}$, corresponding to a low Reynolds number $Re \approx 0.06 \ll 1$.
The recessed sensors were connected to a side hole of diameter $d=1$~mm, while the smallest commercially available flush sensors (diaphragm diameter $D=3.8$~mm) were aligned flush with the wall. A planar channel was selected because the flat diaphragm of a flush sensor cannot be properly aligned with a curved wall \cite{James1990,James1990_2,Rothstein1999,Rothstein2001}. All experiments were conducted at 22~$^\circ$C, and viscous heating is negligible: a simple adiabatic estimate gives a temperature rise below $0.1~^\circ$C.

We first measured the pressure difference $\Delta$ in a straight channel of height $h=1$~mm and test length $l_c=30$~mm (Fig.~\ref{fig3}B). The recessed and flush sensors yielded nearly identical pressure drops. The measurements also agree with simulations using the Carreau model (gray curve), which captures the shear-rate dependence of viscosity, whereas both the Newtonian prediction (black line) and the power-law approximation (blue line) substantially deviate from the measured pressure drop. This confirms that shear thinning must be included to obtain the correct reference pressure drop.

Because the Oldroyd-B and FENE-CR models assume constant shear viscosity, their predictions coincide with the Newtonian result and, therefore, cannot reproduce the measured pressure drop, even in a straight channel. The FENE-P model also fails to capture the shear-thinning behavior (Fig.~\ref{figs_rheo_PIB_PB}D in Appendix~\ref{appendix_fluids}). These limitations motivate the use of the modified FENE-CR (mFENE-CR) model, in which the solvent viscosity depends on shear rate to effectively account for the shear-thinning properties of polymer solutions. Although the FENE-CR model is not derived from rigorous statistical mechanics like the FENE-P model, it remains widely used due to its ability to represent viscosity variations and its favorable numerical stability~\cite{Szabo1997, LpezAguilar2016, TamaddonJahromi2016}. Later, we adopt it as a representative constitutive model for qualitative comparison.

Closer inspection of the individual sensor readings shows that the hole pressure effect $\Delta P_h \propto Q^2$ between recessed and flush configurations (Fig.~\ref{fig3}C) agrees with the prediction of Eq.~\eqref{3d_results}. Moreover, its magnitude is comparable to the rheometric measurements of $N_1$, although the quantitative results $\Delta P_h \approx N_1/3$ suggest that the assumptions underlying the theoretical estimate are not perfectly satisfied, which can be considered for future work. Thus, in a straight channel where the upstream and downstream flow fields are identical, the hole pressure effect cancels when forming the pressure difference. Consequently, recessed and flush sensors yield the same pressure drop, i.e., $\Delta P = \Delta_r = \Delta_f$.
The results imply that, for contraction--expansion flows, the hole pressure effect can, in principle, be removed if the inlet and outlet geometries are identical ($A_{\mathrm{en}} = A_{\mathrm{ex}}$) and the downstream length $l_{\mathrm{ex}}$ is sufficient for polymer relaxation. However, Table~\ref{table_summary} shows that for the experiments reported, the resulting values of $De_{\mathrm{ex}} = \lambda Q/(A_{\mathrm{ex}} l_{\mathrm{ex}})$ are not small, implying that the polymer does not fully relax downstream, which suggests that the hole pressure effect was likely present in many earlier experiments.

\subsection{Pressure drop measurements in a smoothly varying channel}

\begin{figure*}[t!]
\centering
\includegraphics[width=\linewidth]{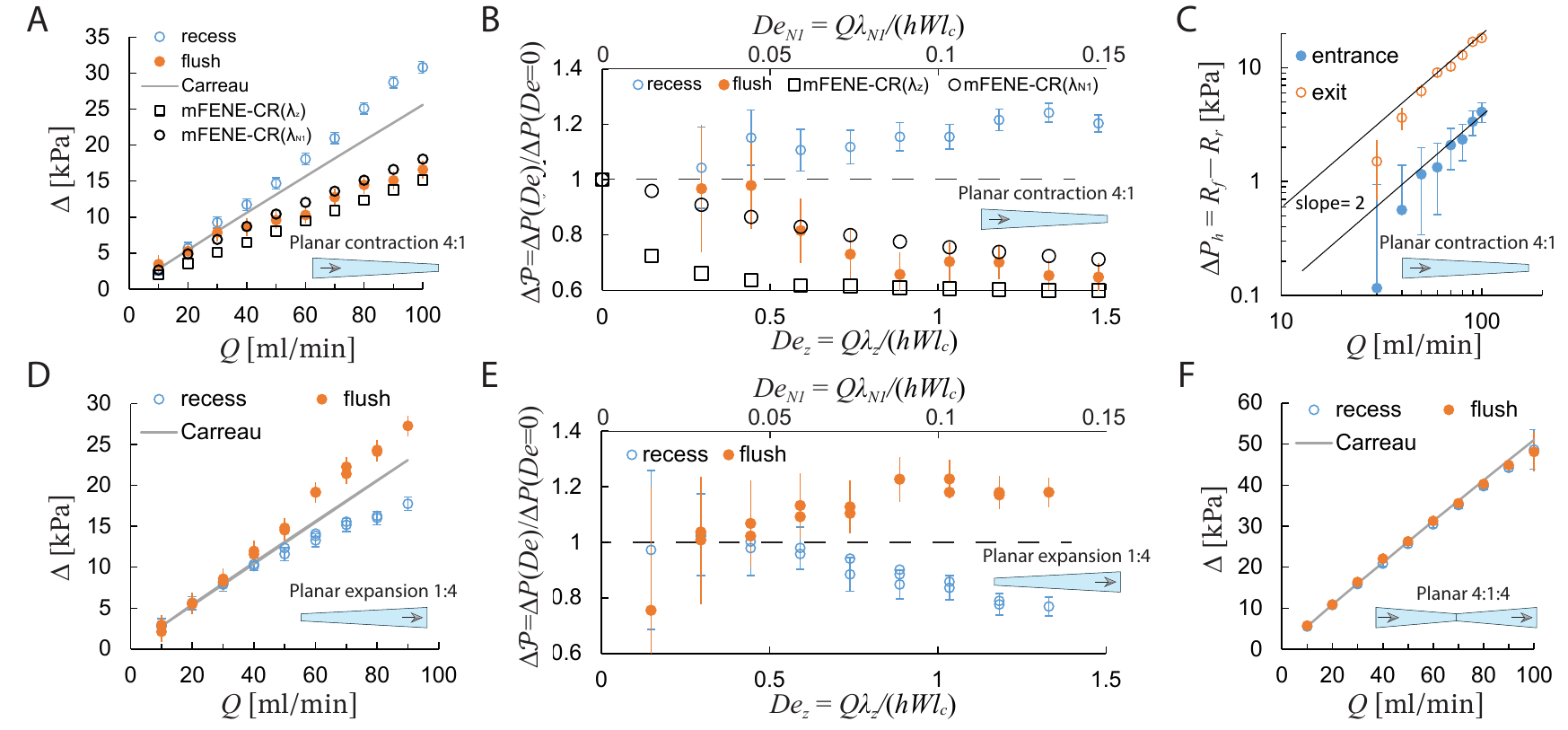}
\caption{Pressure measurements in channels with smoothly varying cross-sectional areas. 
A 4:1 smooth contraction channel $\{H_{\rm en},h,l_c\}=\{4,1,25.6\}$ mm: 
(A) Sensor reading difference $\Delta$, 
(B) normalized pressure drop $\Delta \mathcal{P}$, and 
(C) hole pressure effect difference $\Delta P_h$ as a function of the flow rate $Q$ or Deborah number ($De_z$ or $De_{N1}$). 
A 1:4 smooth expansion channel $\{h,H_{\rm ex},l_c\}=\{1,4,25.6\}$ mm: 
(D) Sensor reading difference $\Delta$ and 
(E) normalized pressure drop $\Delta \mathcal{P}$ as a function of $Q$ or $De_z$. 
For the 4:1:4 smooth contraction--expansion channel $\{H_{\rm en},h,H_{\rm ex},l_c\}=\{4,1,4,55.6\}$ mm: 
(F) Sensor reading difference $\Delta$ as a function of $Q$. In (A) and (B), { black circles for $\lambda_{N1}$ and squares for $\lambda_z$} represent the numerical results of the modified  FENE-CR (mFENE-CR) model obtained from three-dimensional simulations, as detailed in Appendix~\ref{appendix_simulation}.}
\label{fig4}
\end{figure*}
Smooth channels permit theoretical analysis under the lubrication approximation and have recently been a well-studied geometry \cite{Boyko2022,Housiadas2023,Boyko2024,Hinch2024,Mahapatra2025}.
We measured the pressure drop in a smooth $4{:}1$ contraction (geometric details are provided in Table~\ref{tab:geometry}). In this geometry, the recessed and flush sensors yielded markedly different results. As shown in Fig.~\ref{fig4}A, the recessed sensors measured a larger pressure difference $\Delta_r$ than that of a purely viscous fluid, consistent with earlier reports of an “enhanced pressure drop’’ \cite{Boger1987,White1987}. In contrast, the flush sensors reported a pressure drop $\Delta_f=\Delta P$ \emph{smaller} than the viscous reference, in qualitative agreement with theoretical predictions from Oldroyd-B and FENE-type models~\cite{Boyko2022,Housiadas2023,Boyko2024,Hinch2024,Mahapatra2025}.

Three-dimensional numerical simulations under the same conditions (see inelastic fluids in Appendix~\ref{appendix_carreau} and viscoelastic fluids in Appendix~\ref{appendix_simulation}) further support this observation. The modified FENE-CR (mFENE-CR) model, using material parameters obtained independently from rheological tests, quantitatively captured the flush-sensor data for both measurable relaxation times (black circles for $\lambda_{N1}$ and squares for $\lambda_{z}$ in Fig.~\ref{fig4}A), despite the strong disparity between the two values.

This agreement demonstrates that, once the hole pressure effect is removed, FENE-type constitutive modeling can successfully describe the pressure drop of viscoelastic fluids in contraction flows, at least at the level of the simple representative comparison considered here. More detailed and fully self-consistent simulations, incorporating more complete polymer-chain dynamics and rheological responses, remain an important direction for future work.

We next present the normalized pressure drop $\Delta\mathcal{P}$ as a function of the Deborah number $De_z$ and $De_{N1}$ in Fig.~\ref{fig4}B. At low $De_z$ (small $Q$), the response remains nearly Newtonian with $\Delta\mathcal{P}\approx 1$ within sensor sensitivity. As $De_z \gtrsim 0.5$, viscoelastic effects become evident, and for $De_z \gtrsim 1$, the normalized pressure drop approaches a plateau. This behavior is quantitatively captured by the mFENE-CR simulations and is qualitatively consistent with previous numerical studies \cite{Mahapatra2025}. Using $De_{N1}$ based on $\lambda_{N1}$ yields seemingly improved agreement but limits the Deborah number to a much smaller range $De_{N1}\lesssim 0.15$ compared to $De_{z}\lesssim1.5$. Although the choice of relaxation time does not alter the qualitative agreement between experiment and simulation, it complicates a strict quantitative comparison due to the ongoing debate of the choice of relaxation time \cite{James2021}, which is associated with mapping the dimensional $\Delta P$--$Q$ relation into dimensionless $\Delta\mathcal{P}$--$De$ plots of experimental results. Examination of the hole pressure effect $\Delta P_h$ (Fig.~\ref{fig4}C) shows an $\propto Q^2$ scaling, as also observed for the straight channel (Fig.~\ref{fig3}C). In a contraction, the upstream and downstream wall shear rates differ, leading to unequal $\Delta P_{h,\mathrm{en}}$ and $\Delta P_{h,\mathrm{ex}}$ that reflect the geometric area change. These results clarify the fundamentally different behaviors of recessed and flush sensors in channels with monotonically varying cross sections.

Next, we measured the pressure drop in a $1{:}4$ smooth expansion. As expected, the trends in sensor reading difference $\Delta$ (Fig.~\ref{fig4}D) and the normalized pressure drop $\Delta\mathcal{P}$ (Fig.~\ref{fig4}E) are opposite to those observed in the contraction, consistent with previous theoretical predictions \cite{Boyko2022,Hinch2024}. Two distinct transition points again occur, namely one near $De_z \approx 0.5$ and another for $De_z \gtrsim 1$. In this case, the flush sensors capture the true behavior with $\Delta\mathcal{P} > 1$, whereas the recessed sensors misleadingly suggest $\Delta\mathcal{P} < 1$. It is worth noting that the trends shown in Fig.~\ref{fig4}B and \ref{fig4}E are not perfectly symmetric for the same type of sensor, which may originate from the out-of-equilibrium elastic shear stresses that are not symmetric due to the nonlinear elasticity.

In addition, we examined a $4{:}1{:}4$ smooth contraction--expansion channel. In this symmetric geometry, both recessed and flush sensors reported pressure drops indistinguishable from the viscous reference (Fig.~\ref{fig4}F). This finding indicates that the polymer stretch states at the entrance and exit are sufficiently similar that the measured pressure drops fall within the sensor accuracy. Two caveats, however, should be noted. At $Q=90$~ml/min ($De_z\approx0.6$), a slight decrease with $\Delta\mathcal{P}<1$ was detected, and at $Q=100$~ml/min ($De_z\approx0.7$), clear pressure fluctuations appeared (Fig.~\ref{S2}), signaling the onset of flow instability. Under such unstable conditions, the resulting pressure-drop data cannot be interpreted unambiguously.

\subsection{Pressure drop measurements in abruptly varying channels}
\begin{figure}
\centering
\includegraphics[width=\textwidth]{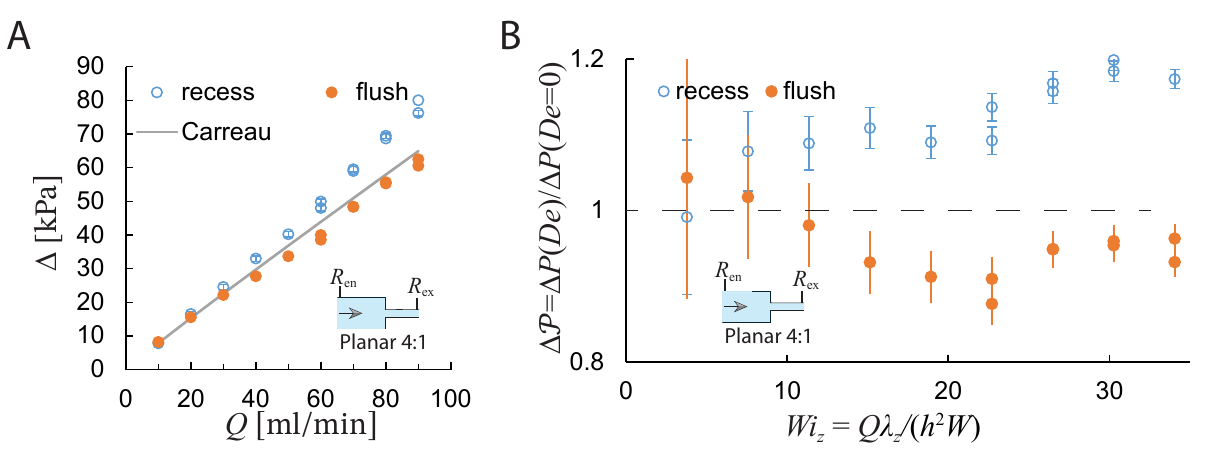}
\caption{Pressure measurements in {planar} abrupt contraction ($4:1$) channels $\{H_{\rm en},h,H_{\rm ex}\}=\{4,1,1\}$ mm. (A) Sensor reading difference $\Delta$ and 
(B) normalized pressure drop $\mathcal{P}$ as a function of the flow rate $Q$ or Weissenberg number ($Wi_z$).}
\label{S3}
\end{figure}
For an abrupt contraction flow, the pressure difference readings $\Delta$ (Fig.~\ref{S3}A) are similar to the smooth contraction case, i.e., the flush sensor records a smaller pressure drop, while the recessed sensor gives an artificially larger value. Thus, abrupt contraction flows also exhibit $\Delta\mathcal{P}<1$ (Fig.~\ref{S3}B). However, the magnitude of the pressure reduction is less than the smooth case, indicating greater dissipation in the abrupt configuration, i.e.,  $\Delta\mathcal{P}_{\mathrm{smooth}} < \Delta\mathcal{P}_{\mathrm{abrupt}} <1$.

\begin{figure*}[t!]
\centering
\includegraphics[width=\linewidth]{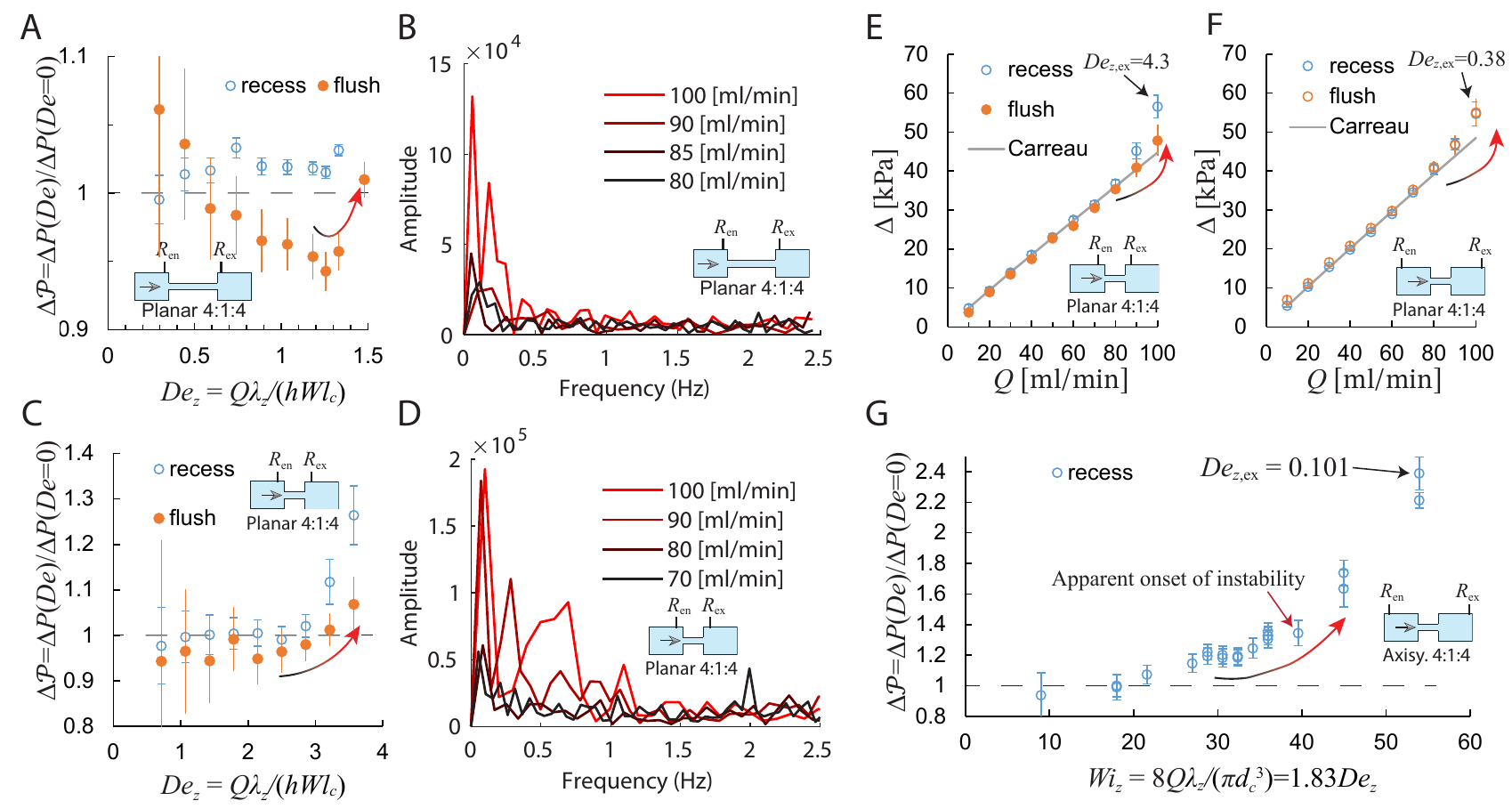}
\caption{Pressure measurements in abrupt $4{:}1{:}4$ contraction--expansion channels with different geometries and sensor positions. 
For a longer connecting channel $\{H_{\rm en},h,H_{\rm ex},l_c\}=\{4,1,4,25.6\}$ mm, 
(A) normalized pressure drop $\Delta\mathcal{P}$ versus Deborah number ($De_z$), and 
(B) FFT spectrum of the flush-sensor signal. 
For a shorter connecting planar channel $\{4,1,4,10.6\}$ mm, 
(C) normalized pressure drop $\Delta\mathcal{P}$ versus $De_z$,
(D) corresponding FFT spectrum. Pressure difference reading $\Delta$ at (E) $l_{\text{ex}}=l_{\text{en}}= 24.6$ mm  and (F) $l_{\text{ex}}=l_{\text{en}}= 24.6$ mm. (G) $\Delta\mathcal{P}$--$Wi_z$ in axisymmetric channels with geometry $\{D_{\rm en}, d_c, D_{\rm ex}, l_c\} = \{7, 1.75, 7, 1.6\}$ mm, where $D$ and $d$ denote diameters at different positions.
}
\label{fig5}
\end{figure*}

We next measured the pressure drop in the widely studied abrupt contraction--expansion geometry, including cases with intermediate connecting lengths of varying size. For sufficiently long connectors, as shown in Fig.~\ref{S4}A, both recessed and flush sensors reported $\Delta$ that closely match the Carreau fluid baseline results over the range $Q=10$–$40$~ml/min ($De_z\lesssim0.75$). At higher flow rates, however, the two sensor types begin to deviate from the baseline prediction.  
The normalized pressure drop $\Delta\mathcal{P}$ (Fig.~\ref{fig5}A) shows that the flush sensors capture a decrease in pressure drop relative to the viscous baseline, consistent with theoretical expectations~\cite{Szabo1997,Alves2003}. In contrast, the recessed sensors again produce the opposite trend. Two distinct transition points, near $De_z\approx 0.5$ and $De_z\gtrsim 1$, are evident, as in the smooth contraction and expansion cases.  

A notable experimental feature of our measurements in this geometry with an abrupt change in cross-sectional area is that, at $Q=100$~ml/min, the flush pressure sensors measured $\Delta\mathcal{P}>1$, in contrast to the trend at lower flow rates. To investigate this apparent anomaly, we performed spectral analyses of the pressure signals in the range $Q=80$–$100$~ml/min (Fig.~\ref{fig5}B). The results showed strong pressure fluctuations at $Q=100$~ml/min, implying the onset of flow instabilities. Hence, we attribute the observed increase in $\Delta\mathcal{P}>1$ at high flow rates to these instability-driven fluctuations rather than to a true steady viscoelastic pressure drop. 

For the contraction--expansion geometry with a shorter connector, the recessed and flush pressure difference readings $\Delta$ showed little deviation from the Carreau fluid baseline over the range $Q=10$–$80$~ml/min (Fig.\ref{S4}B). However, at higher flow rates of $Q=90$ and $100$~ml/min, both kinds of sensors recorded measurements such that $\Delta\mathcal{P}>1$ (Fig.~\ref{fig5}C). Spectral analyses of the pressure signals for the flush-mounted sensor in the range $Q=70$–$100$~ml/min revealed pronounced pressure fluctuations at $Q=90$ and $100$~ml/min (Fig.~\ref{fig5}D), again implying the presence of flow instabilities. 

We further examined the effect of positioning the sensors farther from the contraction--expansion region. The results show that placing the sensors sufficiently downstream avoids the hole pressure effect compared to closer mounting (Fig.~\ref{fig5}E), because the polymer has reached equilibrium at that location, where $De_{z,\text{ex}} = Q\lambda_z /(W H_{\text{ex}} l_{\text{ex}}) = 0.38$ (Fig.~\ref{fig5}F). However, this placement also masks the slight $\Delta\mathcal{P} < 1$ behavior in contraction--expansion flow (Fig.~\ref{fig5}C). Consistent trends were observed when the contraction ratio $H/h$ was increased from 4 to 8, and the connector length was reduced to 1.6 mm: $\Delta\mathcal{P} < 1$ for stable flow and $\Delta\mathcal{P} > 1$ for unstable flow, in the absence of the hole pressure effect (Fig.~\ref{S5}).

In addition to the planar abrupt contraction--expansion geometry, we reexamined the axisymmetric $4{:}1{:}4$ contraction--expansion flow previously studied by Rothstein and McKinley~\cite{Rothstein1999}. The results (Fig.~\ref{fig5}G and Fig.~\ref{S6}) reveal a key distinction from the planar case: a regime with $\Delta\mathcal{P} > 1$ appears at $De_z \approx 10$ ($Wi_z \approx 20$), while spectral analysis shows that instability does not occur until $De_z \approx 20$ ($Wi_z \approx 40$). This indicates that enhanced pressure drops can arise under steady, axisymmetric flow conditions, consistent with earlier observations~\cite{Rothstein1999,Rothstein2001} at $Wi_z\approx12$ and $\approx55$.
Recent simulations using a FENE-P model for planar hyperbolic contraction--expansion flows \cite{Zografos2020} reported a slight decrease in pressure drop ($\Delta\mathcal{P}<1$) for $Wi \le 10$ but a significant increase ($\Delta\mathcal{P}>1$) for $10 \le Wi \le 20$. Although the geometry and rheological parameters differ from the present study, these results suggest that FENE-type simulations may achieve qualitative consistency with both our measurements and prior axisymmetric studies~\cite{Rothstein1999,Rothstein2001} when $Wi_z$ is used instead of $Wi_{N1}$. {Additionally, Koppol et al. \cite{KOPPOL2009} successfully predicted the experimentally observed extra pressure drop by incorporating more detailed microscopic chain dynamics into their simulations. This approach is distinct from conventional FENE-type dumbbell models, although it comes at a substantially higher computational cost.} More geometrically and rheologically consistent simulations will be required for direct quantitative comparison. 

\section{Discussion}

In this work, we examine the discrepancy between theoretical predictions and experimental measurements of the pressure drop--flow rate ($\Delta P-Q$) behavior of dilute viscoelastic polymer solutions flowing through variable-cross-section channels under low-Reynolds-number conditions.  Several factors may contribute to this mismatch: (i) misinterpretation of pressure measurements in piston- or compressed-gas-driven flow systems~\cite{Nigen2002,James2021,James2023}; (ii) the presence of the hole pressure effect, which is particularly significant in contraction or expansion channel~\cite{Rodd2010,Sousa2009}; (iii) insufficient characterization of flow stability, potentially leading to comparisons between unstable experiments and stable numerical simulations; and (iv) inconsistencies in the reported $De$ or $Wi$ ranges arising from different choices of the relaxation time~\cite{Rothstein1999,Sousa2009,Sousa2011b}.

To analyze these issues, we prepared a dilute viscoelastic PIB--PB solution and characterized its rheology through SAOS, CaBER, and steady-shear measurements. These measurements identify that $\lambda_z \approx \lambda_{\mathrm{caber}} \gg \lambda_{N1}$, consistent with existing studies~\cite{Rothstein1999,Sousa2009,Gaillard2025}. Thus, both $\lambda_z$ and $\lambda_{N1}$ were used in our numerical simulations and non-dimensional analysis. Using flush and recessed pressure sensors, we performed systematic measurements in various geometries. These measurements verify the presence of the hole pressure effect and enable qualitative agreements with theoretical predictions across all geometries.
Our measurements show that in slowly varying planar contractions, flush-mounted sensors capture the expected reduction in pressure drop ($\Delta\mathcal{P}<1$), in agreement with simulations, whereas recessed sensors give opposite and misleading trends. These findings, demonstrated here for planar flows, are expected to extend to axisymmetric contractions as well \cite{James1990}.

For contraction--expansion geometries, the response is more nuanced. In smooth channels, similar polymer stretching at the inlet and outlet yields $\Delta\mathcal{P}\approx 1$. In abrupt planar contraction--expansion flows, flush sensors detect a slight decrease ($\Delta\mathcal{P}<1$), consistent with prior simulations. We further show that connector length and sensor placement determine whether $\Delta\mathcal{P}<1$ or $\Delta\mathcal{P}=1$, and that sensors located far downstream largely suppress hole-pressure effects. In axisymmetric flows, our results agree with Rothstein and McKinley \cite{Rothstein1999}: enhanced pressure drops arise at $Wi_z \gtrsim 10$--20, with strong instabilities near $Wi_z \approx 40$--50, qualitatively supported by recent simulations reporting $\Delta\mathcal{P}<1$ for $Wi\le 10$ and $\Delta\mathcal{P}>1$ for $10\le Wi\le 20$, though the geometrical and rheological parameters are different\cite{Zografos2020}. 

In addition, we performed similar systematic experiments using another dilute viscoelastic polymer solution (Appendix~\ref{appendix_another_fluid}), with different polymer molecular weight and solvent composition. The qualitative trends (Fig.~\ref{SI_add_128PIB}B-E) observed are consistent with those reported in the main manuscript for each geometrical configuration. The quantitative comparison for the smooth contraction again shows good agreement. 

Despite the qualitative agreement observed in our experiments, several important questions remain open. In planar flows, the robust trend that $\Delta\mathcal{P}<1$ at low $De$ or $Wi$, and that the $\Delta\mathcal{P}>1$ are accompanied by pressure fluctuations, points to the need for additional flow-visualization studies to more clearly delineate the transition from steady to unsteady flow. Although our axisymmetric results are consistent with earlier reports~\cite{Rothstein1999}, the emergence of a stable $\Delta\mathcal{P}>1$ regime warrants closer examination {for simple dumbbell models, even though Koppol et al. \cite{KOPPOL2009} have shown the trends based on segment-FENE models by considering polymer microdynamics.} In addition, flat-diaphragm pressure sensors introduce hole pressure effects on curved walls, complicating measurements in axisymmetric contractions and expansions.

Moreover, for the constitutive models we considered here, our results demonstrate that no single relaxation time can simultaneously capture both steady shear and extensional rheological responses.  Consequently, although theoretical and numerical models reproduce the qualitative trends, achieving quantitative agreement remains challenging. For example, the choice of relaxation time influences not only the constitutive behavior but also the key non-dimensional parameters ($Wi$, $De$), highlighting the need for caution when comparing experiments with simulations or when interpreting relaxation times across different studies. {The selection of an appropriate relaxation time, therefore, remains an open and actively debated problem in the literature.} This also highlights the need for constitutive models containing more realistic microscopic features of polymer solutions~\cite{BoykoStone2024Perspective}. Further work on these issues is required for complex flows involving both shear and extensional deformations.

Overall, our work provides a detailed experimental investigation, supplemented with simulations, and a discussion of the long-standing claimed pressure drop versus flow rate discrepancy, taking into account factors such as pressure measurement, definitions of dimensionless parameters, and the identification of flow instabilities. Our experiments offer new perspectives and understanding of the discrepancies reported in the literature (as summarized in Table~\ref{tab:exp_summary} in Appendix~\ref{appendix_summary_status_discrepancy}), and suggest that the experimental results are qualitatively consistent with dumbbell-model-based theoretical predictions, without indicating any severe discrepancy. We acknowledge that there remains work to be done.
The remaining issues include, for example, the lack of experimental measurements capturing the regime of $\Delta \mathcal{P} < 1$ in axisymmetric contraction--expansion flows using flush-mounted sensors.
While we have explored a range of geometries and dilute viscoelastic polymer solutions, 
there is value in the future to explore a broader class of viscoelastic fluids, larger Hencky strains, and simultaneous flow field measurements 
for more comprehensive validation.

\section*{acknowledgments}

We thank Gareth McKinley and John Hinch for helpful discussions and.
Dimitrios Fraggedakis for correcting our description of the streamlines coordinate analysis of the hole pressure error. N.H., J.H., and H.A.S. acknowledge the support from grant no.~CBET-2246791 from the United States
National Science Foundation (NSF).
E.B. acknowledges the support by grant no.~2022688 from the US-Israel Binational Science
Foundation (BSF).
N.H. acknowledges the Postdoctoral Fellowship from Zhejiang University.
J.H. acknowledges the Kwanjeong Educational Foundation Graduate Fellowship for financial support. E.B. acknowledges the support from the Israeli Council for Higher Education Yigal Alon Fellowship. We used ChatGPT to check spelling and receive suggestions for grammar.

\startappendices
\section{Literature summary of experiments reporting \texorpdfstring{$\Delta \mathcal{P}>1$}{Delta P>1} and simulations reporting \texorpdfstring{$\Delta \mathcal{P}<1$}{Delta P<1} using dumbbell models}

Table~\ref{table_summary} summarizes previous experimental studies on contraction (C) and contraction--expansion (C--E) flows that reported an apparent enhancement of the normalized pressure drop, i.e., $\Delta\mathcal{P}>1$. In this table, ``n.a.'' denotes ``not available.'' For the fluid classification, ``Boger'' refers to fluids with nearly constant shear viscosity, appearing approximately flat on a logarithmic viscosity--shear-rate plot, whereas ``S.T.'' denotes shear-thinning behavior. For the reported relaxation time, ``$G'=G''$'' indicates that the relaxation time was determined from the crossover point of the storage and loss moduli in oscillatory rheology. For the flow configuration, ``Axisy.'' denotes an axisymmetric geometry; otherwise, the geometry is planar. Unless otherwise specified, the default configuration is an abrupt contraction--expansion geometry. For the pressure-measurement method, ``Diff.'' denotes a recessed differential pressure sensor, ``Two'' denotes two independent recessed pressure sensors, ``One'' denotes a single recessed pressure sensor, ``Air'' denotes the use of compressed air to drive the flow, ``r'' denotes a recessed sensor configuration, and ``f$^*$'' denotes a flush-mounted flat diaphragm sensor embedded in a curved wall. The quantity $Wi_{\mathrm{onset}}=\lambda Q/(Ah)$ is the onset Weissenberg number at which an apparent $\Delta\mathcal{P}>1$ is observed, while $Wi_{\mathrm{pl.}}$ denotes the onset of the plateau in $\Delta\mathcal{P}$. The maximum value of $De_{\mathrm{ex}}=\lambda Q/(A l_{\mathrm{ex}})$, where $A$ is the cross-sectional area, characterizes polymer relaxation after the fluid exits the contraction--expansion region and before it reaches the downstream pressure sensor. When $De_{\mathrm{ex}}\geq \mathcal{O}(1)$, polymer stresses are not expected to fully relax over this downstream distance, indicating that, from a theoretical standpoint, the hole pressure effect cannot be eliminated in such C--E measurements under steady flow conditions.

\begin{table*}[h]
\centering
\caption{Summary of contraction (C) or contraction--expansion (C-E) flow experiments reporting $\Delta\mathcal{P}>1$.}
\label{table_summary}
\resizebox{\textwidth}{!}{
\begin{tabular}{llllllllllllll}
\hline
Ref. & Fluid (concentration) & $h$ or $2r$ [mm] & $l$ [mm] & Ratio & $\lambda$ [s] & Re & De & Wi & Configuration & Method & $Wi_\mathrm{onset}$ & $Wi_\mathrm{pl.}$ & max $De_\mathrm{ex}$ \\
\hline
James 1980 \cite{James1980} & PEO Boger (dilute) & 0.12 & 1.778 & $\approx 36$ & n.a. & $\approx 20$--200 & $\approx 0.1$--1 & $\approx 1.6$--16 & Axisy. C-E (60$^\circ$ conical) & Diff. (r) & n.a. & n.a. & n.a. \\
James 1982 \cite{James1982} & PEO Boger (dilute) & 0.132 & 0.25 & n.a. & n.a. & $\approx 26.4$--132 & $\approx 0.47$--3.17 & $\approx 0.9$--6 & Axisy. C-E (Fillet) & Diff. (r) & n.a. & n.a. & n.a. \\
{Binding 1988 \cite{Binding1988} }& PAM Boger (dilute) & h=2.2, 2r=4.8 & no exit channel & 14 & n.a. & $\le0.3$& n.a. &n.a. &C and Axisy. C (without exit channel) & Four (f) & n.a. & n.a. &n.a.\\
{ Boger 1990 \cite{Boger1990}}& PAM Boger (dilute) & 0.8,1,2.9& n.a. & 4,22& 0.28 & $<0.1$ & n.a. & 14--280 & Axisy. C & n.a.&n.a.&n.a.&n.a. \\
Cartalos 1992 \cite{Cartalos1992} & salted HPAM Boger (dilute) & 1.22 & 1.0858 & 16.4 & n.a. & $\approx 0.015$--10.4 & $\approx 0.45$--236 & $\approx 0.4$--210 & Axisy. C-E & Diff. (r) & n.a. & n.a. & n.a. \\
Rothstein 1999 \cite{Rothstein1999} & PS-PS Boger (dilute) & 6.35 & 12.7 & 4 & 3.24 & $<0.01$ & $\approx 0.04$--4 & $\approx 0.08$--8 & Axisy. C-E & Two (f$^*$) & 12 & n.a. & 0.4 \\
Rothstein 2001 \cite{Rothstein2001} & PS-PS Boger (dilute) & 6.35,3.175,1.5875 & 3.175 & 2,4,8 & 3.24 & $<0.02$ & $\approx 0.05$--20 & $\approx 0.1$--10 & Axisy. C-E & Two (f$^*$) & 9--22 & n.a. & 0.15 \\
Nigen 2002 \cite{Nigen2002} & PAM Boger (unknown) & 8,4,2,1 & n.a. & 4,8,16,32 & n.a. & $<0.15$ & n.a. & n.a. & Axisy. C and planar C & Air & n.a. & n.a. & n.a. \\
Rodd 2005 \cite{Rodd2005} & PEO Boger and S.T. (dilute to semi-dilute) & 0.025 & 0.1 & 16 & 0.7--4.4$\times 10^{-3}$ & 0.44--64 & 68.5 & 274 & C-E & Diff. (r) & 10--50 & n.a. & 0.37 \\
Groisman 2004 \cite{Groisman2004} &  PAA (unknown $\eta(\dot{\gamma})$ and $c$) & 0.037 & 0.22 & $\approx 10$ & 0.013 & <1.6 & $\approx 0.16$--1.6 & $\approx 0.95$--9.5 & C-E (Series of triangle) & n.a. & 2 & 5.2 & n.a. \\
Rodd 2007 \cite{Rodd2007} & PEO Boger and S.T. (dilute to semi-dilute) & 0.025 & 0.1 & 16 & 3.3--20.8$\times 10^{-3}$ & 0.03--11.5 & 0.1--10.5 & 0.4--42 & C-E & Diff. (r) & 10 & 25 & 0.30 \\
Miller 2009 \cite{Miller2009} & PEO S.T. (close to $c^*$) & 0.026 & 0.1 & 15.38 & $\approx 2$ & 0.03--10 & 20 & 0--80 & C-E & Diff. (r) & 15 & n.a. & 0.4 \\
Sousa 2009 \cite{Sousa2009} & PAA Boger (dilute) & 2,3,6,10 & n.a. & 2.4,4,8,12 & 4 & 0.02--4 & n.a. & 0--70 & Square C & Diff. (r) & 0 & n.a. & n.a.\\
Sousa 2010 \cite{Sousa2010} & PEO S.T. (dilute to semi-dilute) & 0.02--0.05 & 0.128--0.382 & 10--20 & 2.5--74$\times 10^{-3}$ & $<30$ & $\approx 40$ & $\approx 200$ & C-E (series of hyperbolic) & Diff. (r) & 10--20 & n.a. & n.a. \\
Campo 2011 \cite{CampoDeao2011} & salted PAA S.T. (dilute) & 0.054 & 0.128 & 7.4 & 4--30$\times 10^{-3}$ & $<500$ & $<10$ & $<30$ & C-E (hyperbolic) & Diff. (r) & 10--15 & n.a. & 1.54 \\
Li 2011 \cite{Li2011} & PEO S.T. (semi-dilute) & 0.102 & 20 & 8 & 6--17$\times 10^{-3}$ & $<5$ & $<0.6$ & $<120$ & C-E & Two (r) & 10 & n.a. & 0.60 \\
Gulati 2008 \cite{Gulati2008} & DNA S.T. (semi-dilute) & 0.1 & n.a. & 2 & 6.8 & $<0.098$ & n.a. & $<629$ & C-E & One (r) & 100 & n.a. & n.a. \\
Rodd 2010 \cite{Rodd2010} & PEO S.T. (semi-dilute) & 0.025 & 0.05--0.4 & 16 & 1.5--3$\times 10^{-3}$ & 0--45 & $<24$ & 0--373 & C-E & Two (r) & 25--50 & n.a. & 3.11 \\
Sousa 2011 \cite{Sousa2011_2} & PAA S.T. (dilute) & 0.054,0.02 & 0.128,0.382 & 7.4,20 & 0.038 & n.a. & $<20$ & $<433$ & C-E (hyperbolic) & Diff. (r) & n.a. & n.a. & 3.09 \\
Lanzaro 2011 \cite{Lanzaro2011} & PAA S.T. (semi-dilute) & 0.2,0.1,0.025 & 20 & 4,8,16 & 2.2--10$\times 10^{-3}$ & 0.02--4 & $<2$ & $<130$ & Axisy. C-E & Diff. (r) & 10 & 150 & 1.20 \\
Perez 2015 \cite{PrezCamacho2015} & PAM Boger (unknown) & 22.5,11.25,7.5,5.62,4.5 & 5 & 2,4,6,8,10 & 0.068 & $<0.1$ & $<20$ & $<4$ & Axisy. C-E & Two (r) & 0.6,1.2,1.4,1.6 & n.a. & n.a. \\
James 2023  \cite{James2023} & PEO S.T. (concentrated) & 3 & n.a. & 4.5 & 2--28 ($G'=G''$) & $<0.1$ & n.a. & $<3000$ & Axisy. C (hyperbolic) & Gas & 47--264 & n.a. & n.a. \\
\hline

\end{tabular}
}
\end{table*}

Table.~\ref{table_simulation_summary} summarizes the representative simulation results based on a continuum model, such as Oldroyd-B, FENE, or PTT models. More complicated models incorporating additional terms \cite{LpezAguilar2016,TamaddonJahromi2016} or micro-dynamics \cite{KOPPOL2009} are not listed here.

\begin{table*}[h]
\centering
\caption{Summary of numerical studies on viscoelastic fluid flows in contraction and contraction--expansion channels using dumbbell models.
}
\resizebox{\textwidth}{!}{%
\begin{tabular}{lccccc}
\hline
Ref & Model& Assumption & Geometry & $Wi$ range & Main results \\
\hline
Keiller 1993~\cite{Keiller1993} & Oldroyd-B, FENE-CR & creeping flow& axisymmetric abrupt or smooth contraction & 0--64 & Oldroyd-B: $\Delta \mathcal{P} < 1$ and  decrease with $Wi$; FENE-CR ($L^2=100$ or 25): $\Delta \mathcal{P} $ decrease then increase with $Wi$, always $< 1$ \\
Szabo 1997~\cite{Szabo1997} & FENE-CR& creeping flow & axisymmetric tube with a 4:1:4 contraction-expansion & 0--10 & Oldroyd-B: $\Delta \mathcal{P} < 1$ and decrease up to $Wi$=3.2; FENE-CR ($L=5$): decrease up to $Wi$=3.2 and increase until $\Delta \mathcal{P} > 1$ at $Wi$=9.6 \\
Alves 2003~\cite{Alves2003} & Oldroyd-B& creeping flow & 2D planar contraction & 0--3 & $\Delta \mathcal{P} < 1$ and decrease with $Wi$ \\
Aguayo 2008~\cite{Aguayo2008} & Oldroyd-B& creeping flow & 2D axisymmetric and planar contraction 4:1 and contraction--expansion 4:1:4 & 1--6.4 & $\Delta \mathcal{P} < 1$ and decrease with $Wi$ \\
Binding 2006~\cite{Binding2006} & Oldroyd-B& creeping flow & axisymmetric and planar contraction 4:1 and contraction--expansion 4:1:4 & 0--6 & $\Delta \mathcal{P} < 1$ decrease with $Wi$ for contraction and contraction-expansion; increase for expansion \\
{Zografos 2020~\cite{Zografos2020}} & FENE-P, sPPT& creeping flow & planar hyperbolic 7.3 contraction--expansion & 0--20 & $\Delta \mathcal{P}<1$ at $Wi<10$ and $\Delta \mathcal{P}>1$ at $Wi>10$ for FENE-P model\\
{Zografos 2022~\cite{Zografos2022}} & FENE-P& creeping flow & planar 4:1 contraction& 0--40 & $\Delta \mathcal{P}<1$ at $Wi<30$ for FENE-P model\\
\hline
\end{tabular}}
\label{table_simulation_summary}
\end{table*}

\section{Derivation of hole pressure effect based on surface coordinates}
\label{appendix_hole_pressure}
\subsection{Three-dimensional stress divergence in local surface coordinates}

In three dimensions, we introduce a local orthonormal basis
$(\mathbf{s}_1,\mathbf{s}_2,\mathbf{n})$, where $\mathbf{s}_1$ and
$\mathbf{s}_2$ are unit tangent vectors to the local surface formed by neighboring
streamlines, and $\mathbf{n}=\mathbf{s}_1\times\mathbf{s}_2$ is the unit normal to the surface. The
two tangent vectors are chosen along the local principal-curvature
directions, with corresponding signed principal curvatures
$\kappa_1$ and $\kappa_2$.

The derivatives of the basis vectors are then
\begin{equation}
\begin{aligned}
\frac{\partial\mathbf{s}_1}{\partial s_1}
&=-\kappa_1\mathbf{n},
&
\frac{\partial\mathbf{n}}{\partial s_1}
&=\kappa_1\mathbf{s}_1,
&
\frac{\partial\mathbf{s}_2}{\partial s_1}
&=\mathbf{0},
\\
\frac{\partial\mathbf{s}_2}{\partial s_2}
&=-\kappa_2\mathbf{n},
&
\frac{\partial\mathbf{n}}{\partial s_2}
&=\kappa_2\mathbf{s}_2,
&
\frac{\partial\mathbf{s}_1}{\partial s_2}
&=\mathbf{0},
\end{aligned}
\label{eq:surface_basis_tangent}
\end{equation}
and
\begin{equation}
\frac{\partial\mathbf{s}_1}{\partial n}
=
\frac{\partial\mathbf{s}_2}{\partial n}
=
\frac{\partial\mathbf{n}}{\partial n}
=\mathbf{0}.
\label{eq:surface_basis_normal}
\end{equation}

Let $\boldsymbol{\sigma}$ be the symmetric second-order stress tensor
written in the local surface coordinates:
\begin{equation}
\begin{aligned}
\boldsymbol{\sigma}
={}&
\sigma_{11}\mathbf{s}_1\mathbf{s}_1
+\sigma_{12}\mathbf{s}_1\mathbf{s}_2
+\sigma_{1n}\mathbf{s}_1\mathbf{n}
\\
&+
\sigma_{12}\mathbf{s}_2\mathbf{s}_1
+\sigma_{22}\mathbf{s}_2\mathbf{s}_2
+\sigma_{2n}\mathbf{s}_2\mathbf{n}
\\
&+
\sigma_{1n}\mathbf{n}\mathbf{s}_1
+\sigma_{2n}\mathbf{n}\mathbf{s}_2
+\sigma_{nn}\mathbf{n}\mathbf{n}.
\end{aligned}
\label{eq:stress_surface_coordinates}
\end{equation}

Using Eqs.~\eqref{eq:surface_basis_tangent} and
\eqref{eq:surface_basis_normal}, the stress divergence is $\left ( \boldsymbol{\nabla}=\mathbf{s}_1
\frac{\partial}{\partial s_1}+\mathbf{s}_2
\frac{\partial}{\partial s_2}+\mathbf{n}
\frac{\partial}{\partial n}
\right )$
\begin{equation}
\begin{aligned}
\boldsymbol{\nabla}\cdot\boldsymbol{\sigma}
={}&
\left[
\frac{\partial\sigma_{11}}{\partial s_1}
+\frac{\partial\sigma_{12}}{\partial s_2}
+\frac{\partial\sigma_{1n}}{\partial n}
+(2\kappa_1+\kappa_2)\sigma_{1n}
\right]\mathbf{s}_1
\\
&+
\left[
\frac{\partial\sigma_{12}}{\partial s_1}
+\frac{\partial\sigma_{22}}{\partial s_2}
+\frac{\partial\sigma_{2n}}{\partial n}
+(\kappa_1+2\kappa_2)\sigma_{2n}
\right]\mathbf{s}_2
\\
&+
\left[
\frac{\partial\sigma_{1n}}{\partial s_1}
+\frac{\partial\sigma_{2n}}{\partial s_2}
+\frac{\partial\sigma_{nn}}{\partial n}
+\kappa_1(\sigma_{nn}-\sigma_{11})
+\kappa_2(\sigma_{nn}-\sigma_{22})
\right]\mathbf{n}.
\end{aligned}
\label{eq:stress_divergence_surface}
\end{equation}

\subsection{Slow flow past a circular hole}

At the centerline of a circular hole, the local geometry is
axisymmetric. We therefore take the two principal curvatures to be
equal:
\begin{equation}
\kappa_1=\kappa_2=\kappa.
\label{eq:equal_principal_curvatures}
\end{equation}
Here $\mathbf{s}_1$ denotes the principal flow direction,
$\mathbf{s}_2$ denotes the transverse tangential direction, and
$\mathbf{n}$ denotes the local normal direction.
By symmetry, the tangential derivatives vanish locally at the
centerline:
\begin{equation}
\frac{\partial}{\partial s_1}
=
\frac{\partial}{\partial s_2}
=0.
\label{eq:centerline_symmetry}
\end{equation}
For a low-Reynolds-number flow with negligible inertia,
$\boldsymbol{\nabla}\cdot\boldsymbol{\sigma}=\mathbf{0}$.
Equation~\eqref{eq:stress_divergence_surface} therefore reduces to
\begin{equation}
\begin{aligned}
\mathbf{0}
=
\left(
\frac{\partial\sigma_{1n}}{\partial n}
+3\kappa\sigma_{1n}
\right)\mathbf{s}_1
+
\left(
\frac{\partial\sigma_{2n}}{\partial n}
+3\kappa\sigma_{2n}
\right)\mathbf{s}_2
+
\left[
\frac{\partial\sigma_{nn}}{\partial n}
+\kappa(\sigma_{nn}-\sigma_{11})
+\kappa(\sigma_{nn}-\sigma_{22})
\right]\mathbf{n}.
\end{aligned}
\label{eq:circular_hole_divergence}
\end{equation}

We further assume that the local flow is approximately unidirectional
along $\mathbf{s}_1$, such that
\begin{equation}
\sigma_{2n}\simeq0.
\label{eq:unidirectional_flow}
\end{equation}
The $\mathbf{s}_1$-directed momentum equation then gives
\begin{equation}
\frac{\partial\sigma_{1n}}{\partial n}
=
-3\kappa\sigma_{1n},
\label{eq:tangential_balance_circular}
\end{equation}
whereas the normally directed momentum equation gives
\begin{equation}
\frac{\partial\sigma_{nn}}{\partial n}
=
\kappa
\left[
(\sigma_{11}-\sigma_{nn})
-
(\sigma_{nn}-\sigma_{22})
\right].
\label{eq:normal_balance_circular}
\end{equation}

Consequently,
\begin{equation}
\kappa
=
-\frac{\partial_n\sigma_{1n}}{3\sigma_{1n}}
=
\frac{\partial_n\sigma_{nn}}
{(\sigma_{11}-\sigma_{nn})-(\sigma_{nn}-\sigma_{22})}.
\label{eq:curvature_circular}
\end{equation}
Eliminating $\kappa$ and integrating along the normal direction yields
\begin{equation}
\int_h^w
\partial_n\sigma_{nn}\,\mathrm{d}n
=
-\frac{1}{3}
\int_h^w
\frac{
(\sigma_{11}-\sigma_{nn})
-
(\sigma_{nn}-\sigma_{22})
}
{\sigma_{1n}}
\,\mathrm{d}\sigma_{1n},
\label{3d_int}
\end{equation}
where $h$ and $w$ denote the positions at the bottom of the hole and
at the wall, respectively. Eq.~\eqref{3d_int} is equivalent to the three-dimensional result derived in earlier work using a general tensorial description with Christoffel symbols~\cite{Higashitani1972}. Here, we recover the same result through a more geometrically transparent formulation based on local surface coordinates.

Defining the first and second normal-stress differences, respectively, by
\begin{equation}
N_1
\equiv
\sigma_{11}-\sigma_{nn},
\qquad
N_2
\equiv
\sigma_{nn}-\sigma_{22},
\end{equation}
Eq.~\eqref{3d_int} becomes
\begin{equation}
\sigma_{nn,w}-\sigma_{nn,h}
=
-\frac{1}{3}
\int_h^w
\frac{N_1-N_2}{\sigma_{1n}}
\,\mathrm{d}\sigma_{1n}.
\label{eq:hole_pressure_general}
\end{equation}

For an Oldroyd--B fluid in steady simple shear,
\begin{equation}
N_1
=
2\eta_p\lambda\dot{\gamma}^{\,2},
\qquad
N_2=0,
\qquad
\sigma_{1n}
=
(\eta_s+\eta_p)\dot{\gamma}.
\label{eq:oldroyd_b_shear}
\end{equation}
It follows that
\begin{equation}
\mathrm{d}\sigma_{1n}
=
(\eta_s+\eta_p)\,\mathrm{d}\dot{\gamma},
\end{equation}
and hence
\begin{equation}
\frac{N_1}{\sigma_{1n}}\,\mathrm{d}\sigma_{1n}
=
\frac{
2\eta_p\lambda\dot{\gamma}^{\,2}
}{
(\eta_s+\eta_p)\dot{\gamma}
}
(\eta_s+\eta_p)\,\mathrm{d}\dot{\gamma}
=
2\eta_p\lambda\dot{\gamma}\,\mathrm{d}\dot{\gamma}.
\end{equation}
Substitution into Eq.~\eqref{eq:hole_pressure_general} gives
\begin{equation}
\begin{aligned}
\sigma_{nn,w}-\sigma_{nn,h}
=
-\frac{1}{3}
\int_{\dot{\gamma}_h}^{\dot{\gamma}_w}
2\eta_p\lambda\dot{\gamma}\,
\mathrm{d}\dot{\gamma}
=
-\frac{1}{3}
\eta_p\lambda
\left(
\dot{\gamma}_w^{\,2}
-
\dot{\gamma}_h^{\,2}
\right).
\end{aligned}
\label{eq:oldroyd_hole_integral}
\end{equation}
Assuming that the flow vanishes sufficiently deep inside the hole,
$\dot{\gamma}_h=0$, and using
$N_{1,w}=2\eta_p\lambda\dot{\gamma}_w^{\,2}$, we obtain
\begin{equation}
\sigma_{nn,w}-\sigma_{nn,h}
=
-\frac{1}{6}N_{1,w}.
\end{equation}

\section{Viscoelastic fluid preparation and characterization}
\label{appendix_fluids}

\begin{figure}
\centering
\includegraphics[width=
0.8\textwidth]{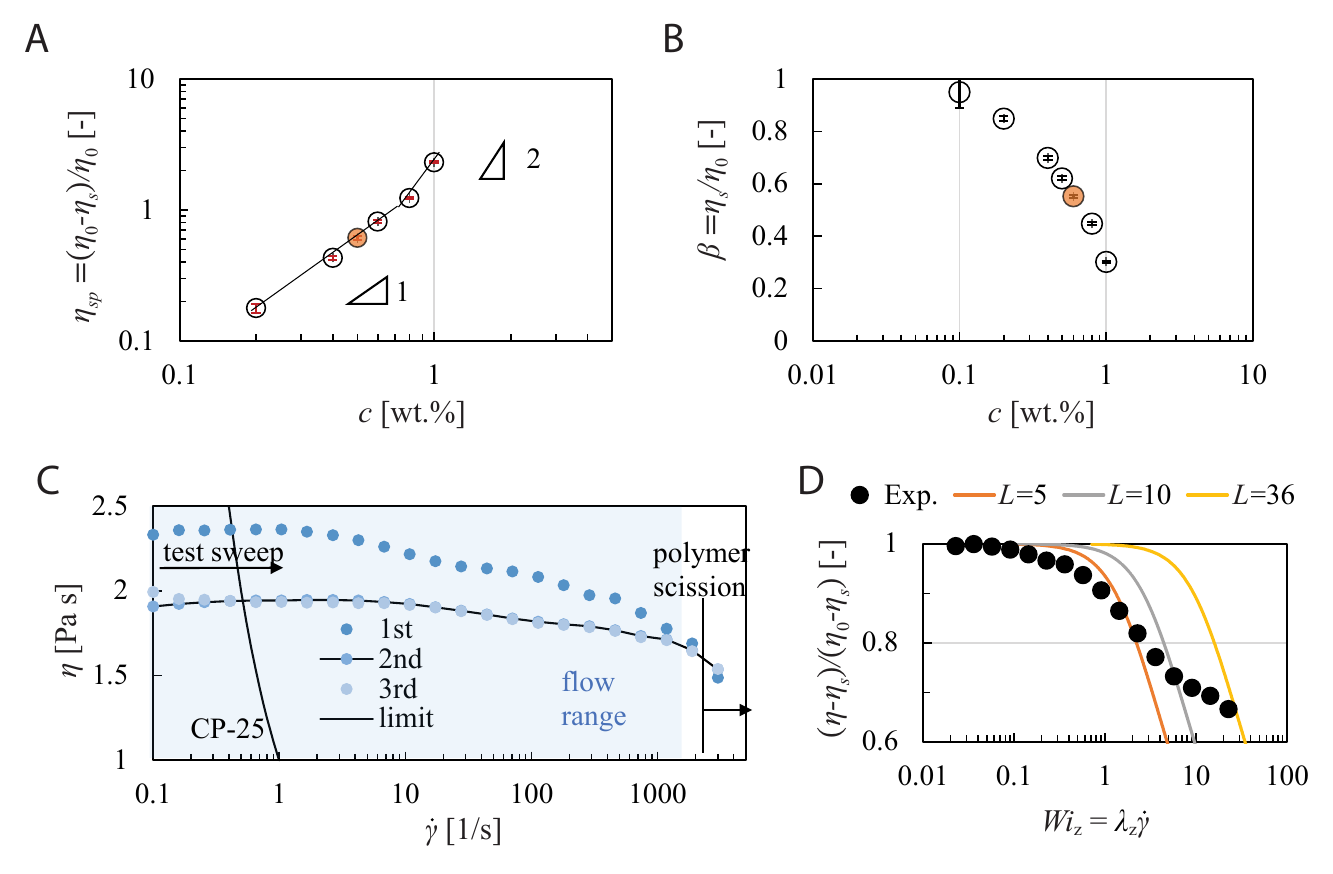}
\caption{Rheology. (A) Specific viscosity $\eta_{\it sp} = (\eta_0 - \eta_s)/\eta_s$ and (B) viscosity ratio $\beta = \eta_s/\eta_0$ as a function of the PIB concentration $c$. (C) Sweep-test results for polymer scission. The labels “1st,” “2nd,” and “3rd” in the legend denote consecutive tests performed on the same sample without reloading. A 25-mm-diameter cone-and-plate geometry was used to access higher shear rates. The orange scatter points in (A) and (B) correspond to $c=0.5$~wt.\%, which was selected as the test fluid in this study. The results in (C) indicate that the onset of noticeable polymer scission occurs at a shear rate of $\approx 2000~\mathrm{s^{-1}}$, higher than the maximum shear rate reached under any of the flow conditions investigated in this work. (D) Comparison between the measured shear viscosity and the FENE-P model predictions for different $L$ values, demonstrating that the FENE-P model cannot capture the shear-thinning behavior observed in this work. Here, $\lambda_z=0.227$ s was adopted for calculating $Wi_z$ for experimental data. The FENE-P prediction is calculated based on the implicit formula from \cite{Yamani2023}.
}
\label{figs_rheo_PIB_PB}
\end{figure}
We first prepared a 3 wt.\% polyisobutylene (PIB) solution by melting and dissolving the PIB polymer (average $M_w=2.8\times10^6$, Sigma-Aldrich) in light mineral oil (Sigma-Aldrich) at 100°C with heating and stirring. Next, we mixed polybutene (PB, average $M_n=2300$, Sigma-Aldrich) and light mineral oil as the solvent, to which the previously prepared PIB-mineral oil solution was added. The final solution, containing {1, 0.8, 0.6, 0.5, 0.4, and 0.2} wt.\% PIB polymer with 40 wt.\% PB and 60 wt.\% light mineral oil as the solvent, was obtained through mechanical stirring followed by vacuum degassing. 

The viscosity $\eta$ and the storage and loss moduli ($G'$ and $G''$) of each solution were measured using a rheometer (Anton Paar 302e) with a cone-and-plate (CP-50) geometry. The measured specific viscosity $\eta_{sp} = (\eta_0 - \eta_s)/\eta_0$ confirmed that solutions with concentrations up to 0.6 wt.\% can be regarded as in the dilute regime (Fig.~\ref{figs_rheo_PIB_PB}A).
Considering the concentration range and the viscoelastic properties, the 0.5 wt.\% PIB solution was selected as the test fluid for all experiments in this work (Fig.~\ref{figs_rheo_PIB_PB}A and Fig.~\ref{figs_rheo_PIB_PB}B).
In order to examine the possibility of polymer scission, a smaller cone-and-plate with diameter of 25 mm (CP-25) geometry was employed, allowing shear rates $\dot{\gamma}$ up to 3000 s$^{-1}$. Repeated shear rate sweep tests from low to high values revealed that the first noticeable effect of polymer scission on viscosity occurred at approximately 2000 s$^{-1}$, which is significantly higher than the maximum shear rate encountered in this study (Fig.~\ref{figs_rheo_PIB_PB}C). Therefore, it can be concluded that polymer scission did not influence the pressure drop measurements within the operating conditions of this study.

\section{Relaxation time measurements and determination}
\label{appendix_relaxation_time}
\subsection{Zimm model and relaxation time \texorpdfstring{$\lambda_z$}{lambda\_z} from dynamic modulus fitting}
\label{appendix_zimm}
The PIB–PB Boger fluid can be described by the Zimm model \cite{More2023,2025Jong}. Specifically, the Zimm model expresses the dynamic moduli of a viscoelastic fluid as
\begin{equation}
\begin{aligned}
   G^{\prime}(\omega)=G_p \frac{\omega \lambda_z \sin [\chi \arctan (\omega \lambda_z)]}{\left[1+(\omega \lambda_z)^2\right]^{\chi / 2}}, \\
   G^{\prime \prime}(\omega)-\eta_s \omega=G_p \frac{\omega \lambda_z \cos [\chi \arctan (\omega \lambda_z)]}{\left[1+(\omega \lambda_z)^2\right]^{\chi / 2}} ,
\end{aligned}
\end{equation}
where $G^{\prime}(\omega)$ and $G^{\prime\prime}(\omega)$ denote, respectively, the storage and loss moduli that describe the elastic and viscous responses of the fluid at angular frequency $\omega$, respectively. Here, $\eta_s$ is the solvent viscosity, $G_p$ is the modulus, $\lambda$ is the relaxation time, and $\chi$ is the Zimm exponent associated with hydrodynamic interactions in dilute polymer solutions ($\chi=0.33$ for a $\theta$-solution).
By fixing $\eta_s = 1.45$~Pa s from experimental measurements and taking $\chi = 0.33$ together with $G_p = (\eta_0 - \eta_s)/\lambda_z$, the fitting yields a relaxation time of $\lambda_z = 0.227$~s for the data shown in Fig.~\ref{fig2}B of the main text.

\subsection{Relaxation time \texorpdfstring{\(\lambda_{\text{caber}}\)}{lambda_caber} measurement based on capillary breakup}
\label{appendix_caber}
\begin{figure}[h]
\centering
\includegraphics[width=0.7\textwidth]{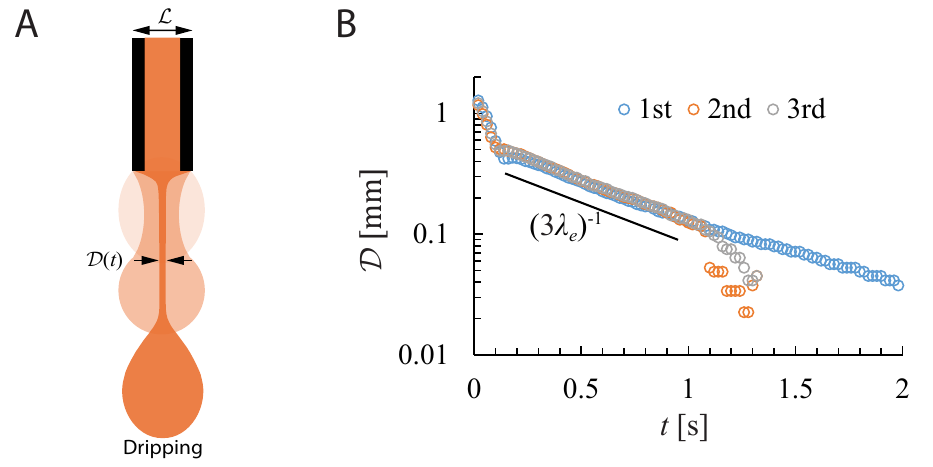}
\caption{Relaxation time measurement based on capillary breakup. (A) Schematic illustration of the dripping process from a nozzle with an outer diameter $\mathcal{L}$, during which the filament diameter $\mathcal{D}$ was monitored.
(B) Temporal evolution of filament diameter $\mathcal{D}(t)$, where the early stage of the elastocapillary balance was fitted using the scaling law $(3\lambda_e)^{-1}$ to extract the apparent relaxation time $\lambda_e = 0.213 \pm 0.006$ s.}
\label{CaBER}
\end{figure}

Following the procedure described in our previous work~\cite{Hu2025}, we employed a dripping method (Fig.~\ref{CaBER}A), using a nozzle with an outer diameter of $\mathcal{L}=2.16$~mm (corresponding to a 12~gauge needle) to minimize pre-stretching effects~\cite{Hu2025}. An apparent relaxation time of $\lambda_{e,\text{caber}} = 0.213 \pm 0.006$~s was obtained by fitting the elastocapillary thinning stage with the scaling law $(3\lambda_{e,\text{caber}})^{-1}$ (Fig.~\ref{CaBER}B). 
The measured elastocapillary number~\cite{Hu2025} was 
$Ec_e = 2\gamma\lambda_{e,\text{caber}}/(\mathcal{D}_0 \eta_p) \approx 15$,
where the surface tension is $\gamma = 29$~mN/m, the initial filament diameter in the breakup process is $\mathcal{D}_0 \approx 1$~mm, and the polymeric viscosity is $\eta_p = 0.804$~Pa s. According to the phase diagram characterizing the dripping measurement method~\cite{Hu2025}, these parameters correspond to a ratio $\lambda_{e,\text{caber}} / \lambda_{\text{caber}} \approx 0.95$, giving the actual relaxation time as 
$\lambda_{\text{caber}} \approx 0.223 \pm 0.006$ s.
This value is consistent with that obtained from oscillatory shear measurements.
{
\subsection{Relaxation time \texorpdfstring{$\lambda_{N1}$}{lambda	extsubscript{N1}} from first normal-stress difference \texorpdfstring{$N_1$}{N1}}
\label{appendix_N1}
Since Fig.~\ref{fig2}A shows that the shear viscosity $\eta$ is approximately constant and the first normal stress difference $N_1$ approximately scales as $N_1 \propto \dot{\gamma}^2$, the Oldroyd-B or FENE-CR constitutive models give
$N_1 = 2 \eta_p \lambda_{N1}\dot{\gamma}^2$.
Using the data in Fig.~\ref{fig2}A and applying a least-squares fit, we obtain an estimated relaxation time of $\lambda_{N1} \approx 0.0228~\mathrm{s}$.

\subsection{Discussion on various relaxation times}
\label{appendix_relaxation_discuss}
Both in the literature and in the present work, three different approaches are commonly used to determine the characteristic relaxation time of dilute polymer solutions. Specifically, $\lambda_z$ is obtained by fitting the modulus data from small-amplitude oscillatory-shear (SAOS) measurements based on the Zimm model, $\lambda_{N1}$ is calculated from the measured first normal-stress difference using $\lambda_{N1} = N_1/(2\eta_p\dot{\gamma}^2)$, and $\lambda_{\mathrm{caber}}$ is determined from extensional-flow measurements using the capillary breakup extensional rheometry (CaBER) method. The specific results obtained in this work, together with classical literature values, are summarized in Table~\ref{table_relaxtion_time}. In general, we find that $\lambda_z \approx \lambda_{\mathrm{caber}} \gg \lambda_{N1}$, which is a trend that is widely observed\cite{Rothstein1999,James2021,Gaillard2025}. This trend can be rationalized by considering the flow conditions and the corresponding microscopic polymer deformation associated with each measurement method.

In SAOS, the material response remains within the linear viscoelastic regime, and the polymer conformations oscillate close to equilibrium. For example, with the fixed strain of $\gamma_0=1\%$ (corresponding the maximum rotating angle $\Omega=\gamma_0\alpha=0.01^{\circ}$, where the cone angle $\alpha=1^{\circ}$) in the cone-and-plate configuration, the shear rate is $\dot{\gamma}=\gamma_0\omega\cos(\omega t)$ and the maximum value of $\dot\gamma\approx1~\mathrm{s}^{-1}$ at $\omega=100$ Hz. Hence, our polymer solution satisfies $Wi=\dot\gamma\lambda\ll1$. Under these conditions, measurements make polymers undergo affine deformation, and the characteristic longest relaxation time $\lambda_z$ can be uniquely determined using the Zimm model (which accounts for hydrodynamic interactions), particularly given that the polymer solution in this study is known to be nearly a $\theta$-solvent.

The CaBER method probes extensional flow, where polymers undergo substantial stretching and are driven far from equilibrium during the elastocapillary-thinning stage under an approximately constant extensional rate. As a result, the relaxation time extracted from CaBER, $\lambda_{\mathrm{caber}}$, also represents the longest polymer relaxation time.

In contrast, although polymers in steady shear flow can also be significantly deformed at high shear rates, linear flexible polymers undergo non-affine deformations with a sequence of stretching, tumbling, and compression cycles \cite{Smith1999}. Consequently, the measured first normal-stress difference $N_1$ reflects a time-averaged effective tensile stress rather than the stress from a steady deformation. This feature provides a qualitative explanation for the long-standing observation that $\lambda_z$ and $\lambda_{\mathrm{caber}}$ are comparable in magnitude and both are substantially larger than $\lambda_{N1}$.}
\begin{table}
    \centering
    \caption{Comparison of relaxation times of dilute polymer solutions from different methods.}
    \begin{threeparttable}
   \begin{tabular}{l l c c c c c}
        \hline
         & solution & $c/c^*$ & $\eta_s$ [Pa s]
         & $^{\text{a}}\lambda_{z}$ [s]
         & $^{\text{c}}\lambda_{\text{caber}}$ [s]
         & $^{\text{b}}\lambda_{N1}$ [s] \\
        \hline
        Current work
        & 2.8 M PIB in PB/oil
        & 0.8
        & 1.45
        & 0.227
        & 0.224
        & 0.0228 \\

        Rothstein et al. \cite{Rothstein1999,Rothstein2001}
        & 2.25 M PS in PS resin
        & 0.24
        & 21
        & 3.24
        & $^{\text{d}}\approx 2$ \cite{Clasen2006}
        & 0.146 \\

        Gaillard et al. \cite{Gaillard2025}
        & 4 M PEO in water
        & 0.37
        & 0.0092
        & /
        & 0.09
        & 0.002 \\

        Anna et al. \cite{Anna2001}
        & 2 M PS in styrene
        & 0.44
        & 34
        & 3.7
        & 3.19
        & / \\
        \hline
        \label{table_relaxtion_time}
    \end{tabular}

    \begin{tablenotes}
        \small
        \item[a] $\lambda_z$ from SAOS data fitting with the Zimm model, considering hydrodynamic interactions between the segments of the polymer chain and the surrounding solvent.
        \item[b] $\lambda_{N1} = N_1 / (2\eta_p \dot{\gamma}^2)$ from shear flow based on the Oldroyd-B model,  and fit with the $N_1$ experimental data.
        \item[c] $\lambda_{\text{caber}}$ extracted from the CaBER method using the Oldroyd-B model or FENE-P model.
        \item[d]
        A similar solution is investigated in \cite{Clasen2006} and and results are shown in Figure 3 of \cite{Clasen2006}, which suggests $\lambda_{\text{caber}}\approx5$--$6$ s for $M_w=2.8$ MDa at the same concentration of 0.025 wt.\% and $\eta_s=51$ Pa s. Since $\lambda \propto \eta_sM_w^{1.5}$ for a $\theta$-solvent, we estimate $\lambda_{\text{caber}}\approx2$ s in \cite{Rothstein1999,Rothstein2001}.
    \end{tablenotes}
    \end{threeparttable}
\end{table}

\section{Experimental setups of pressure measurements and error analysis}
\label{appendix_exp}

\begin{figure}
\centering
\includegraphics[width=\textwidth]{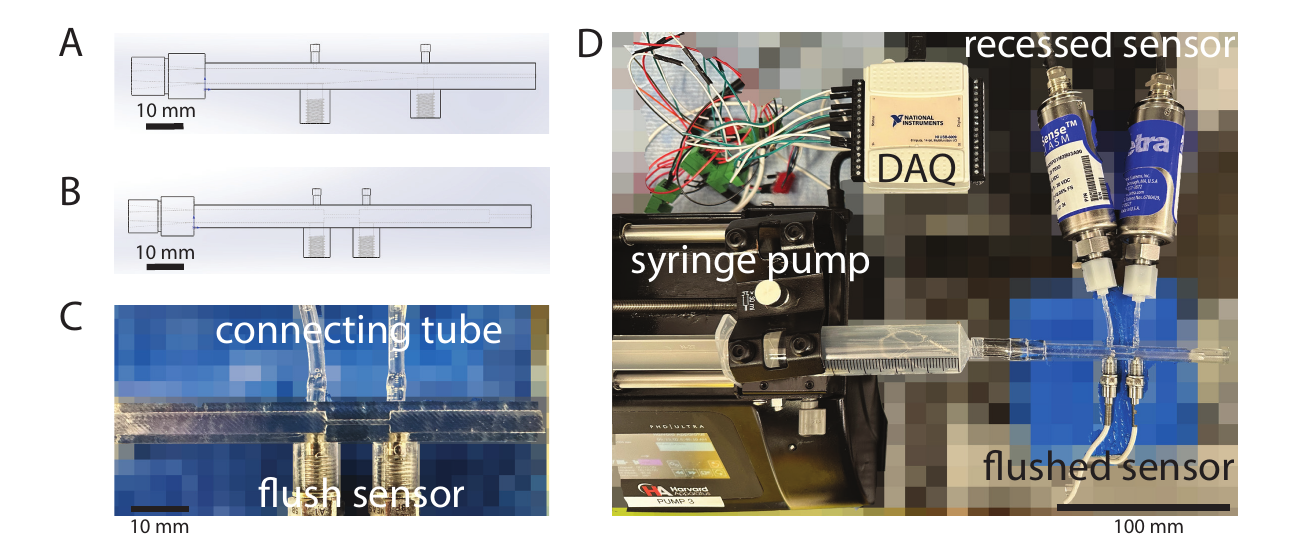}
\caption{Channel design and experimental setup. (A) Geometry of a planar 4:1 smooth contraction channel. (B) Geometry of a planar 4:1:4 abrupt contraction--expansion channel with a short connecting section. (C) Zoom-in view of the sensor mounting of recessed and flush configurations. (D) Schematic of the overall experimental apparatus.}
\label{fig_setup}
\end{figure}

\begin{figure}
\centering
\includegraphics[width=0.8\textwidth]{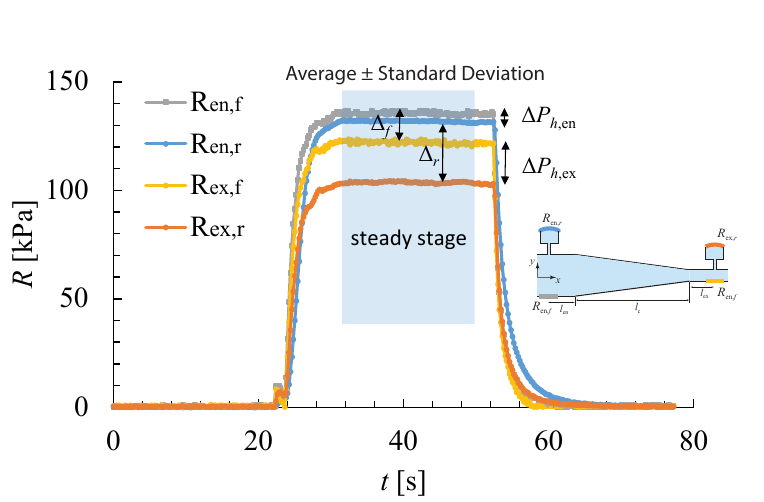}
\caption{Representative pressure recording ($R$) in a 4:1 smooth contraction channel at a flow rate of $Q = 90$ ml/min. The blue-shaded region indicates the interval for data processing, which is used to calculate the mean value and standard deviation of the pressure signal.}
\label{typical_pressure_measure}
\end{figure}

All channels in this work were fabricated using a Formlab 4 3D printer with a printing resolution of 25~$\mu$m. To simultaneously measure pressures with both recess- and flush-mounted sensors, two types of pressure transducers were employed. The two recess-mounted pressure transducers were supplied by Setra (model ASM; measurement range: 0--25~Psi, accuracy: $\sim$200~Pa) with factory calibration. The two flush-mounted pressure transducers were supplied by TE Connectivity (miniature sensor model XP5; measurement range 0--5~bar, accuracy $\sim$600~Pa) with a planar sensing tip diameter of 3.8~mm, and were calibrated against the recess-mounted sensors using fluid flows with glycerol in a 1~mm channel. Voltage signals from all sensors were collected by an NI DAQ 6009 at a sampling frequency of 20~Hz (100~Hz for the experiments with an axisymmetric channel). The recess-mounted sensors were connected to the measurement points via tubing with an inner diameter of 1 mm. The flush-mounted sensors were installed through threaded fittings printed with the flow channels (Fig.~\ref{fig_setup}). We adjusted the angle for the threads to ensure that the sensing surface was nearly coplanar with the channel wall. Previous studies of viscoelastic slit flows using flush-mounted sensors with diaphragms larger than the slit height have shown that, with proper sensor arrangement, the diaphragm size does not affect pressure measurements \cite{Novotny1973, Padmanabhan1994, Han1993}. Consistently, our calibration experiments in straight and contraction--expansion channels showed agreement between flush-mounted and recess-mounted sensors, with the flush diaphragms carefully aligned with the channel wall to minimize geometric deviations. All the geometrical parameters of measured channels are listed in Table~\ref{tab:geometry}.

Prior to each experiment, the working fluid was degassed under vacuum. The fluid was then slowly drawn from the reservoir to fill a syringe (BD, 60~ml), which was mounted on a syringe pump (Harvard PHD Ultra). After initiating data acquisition and recording a baseline voltage (pressure) signal, the syringe pump was operated at a constant flow rate in the range of 10--100~ml/min, with increments of 10~ml/min. For each channel geometry and flow-rate condition, three independent measurements were performed to ensure reproducibility of the results. 
During data processing, only the pressure data corresponding to the plateau region were considered as steady-state results. The mean value and standard deviation of this region were calculated; as an example, see Fig.~\ref{typical_pressure_measure}. For high-flow-rate conditions, the recorded pressure time series were imported into MATLAB for frequency-domain analysis. The reported errors in this work are defined as the larger of (i) the nominal accuracy of the pressure sensors and (ii) the standard deviation of the averaged measurements.

\begin{table*}[htbp]
\centering
\caption{Geometrical details of tested channels (units: mm).}
\label{tab:geometry}
\resizebox{\textwidth}{!}{%
\begin{tabular}{lcccccc}
\hline
Type & $l_c$ ($l_{\it con}$) & $h$ or $d_c$ & $H_{\rm en}$ or $D_{\rm en}$  & $H_{\rm ex}$ or $D_{\rm ex}$ & $l_{\rm en}$ & $l_{\rm ex}$ \\
\hline
Planar straight & 30 & 1 & 1 & 1 & 0 & 0 \\
Planar smooth contraction 4:1 & 25.6 & 1 & 4 & 1 & 2.2 & 2.2 \\
Planar abrupt contraction 4:1 & / & 1 & 4 & 1 & 17.2 & 17.8 \\
Planar smooth expansion 1:4 & 25.6 & 1 & 1 & 4 & 2.2 & 2.2 \\
Planar smooth contraction--expansion 4:1:4 & 55.6 (4.4) & 1 & 4 & 4 & 2.2 & 2.2 \\
Planar abrupt contraction--expansion 4:1:4 & 25.6 & 1 & 4 & 4 & 2.2 & 2.2 \\
Planar abrupt contraction--expansion 4:1:4 & 10.6 & 1 & 4 & 4 & 2.2 & 2.2 \\
Planar abrupt contraction--expansion 4:1:4 far meast. & 10.6 & 1 & 4 & 4 & 24.6 & 24.6 \\
Planar abrupt contraction--expansion 4:0.5:4 far meast. & 1.6 & 0.5 & 4 & 4 & 29.2 & 29.2 \\
Axisymmetric abrupt  contraction--expansion 4:1:4 far meast. & 1.6 &1.75 & 7& 7 & 29.2 & 29.2 \\
\hline
\end{tabular}}
\begin{flushleft}
\footnotesize $l_c$ is the constriction length, $l_{\it con}$ is the straight connecting length, $h$ is the minimum height of the planar channel, $d_c$ is the minimum diameter of the axisymmetric channel, $H_{\rm en}$ and $H_{\rm ex}$ are the entrance and exit heights, respectively, $D_{\rm en}$ and $D_{\rm ex}$ are the entrance and exit diameters, respectively, and $l_{\rm en}$ and $l_{\rm ex}$ are the upstream and downstream sensor distances. The width is $W=10$~mm for the planar geometry.
\end{flushleft}
\end{table*}

\section{Numerical simulations of pressure drop of reference inelastic fluid}
\label{appendix_carreau}
The reference pressure drop $\Delta P(De=0;Q)$ for different channels were obtained from numerical simulations, which were carried out using the laminar flow module in COMSOL Multiphysics. The Carreau fluid is adopted as $\eta(\dot{\gamma}) = \eta_\infty + \left( \eta_0 - \eta_\infty \right) 
[ 1 + \left( \lambda_{ca} \dot{\gamma} \right)^2 ]^{(n-1)/2}$. Parameters obtained from experimental fitting were used: $\eta_0 = 2.284$~Pa s, $\eta_{\infty} = 1$~Pa s, $\lambda_{ca} = 0.4$~s, and $n = 0.919$, which is consistent with the grey curve in Fig.~\ref{fig2}B. { The power-law model adopted is described as $\eta(\dot{\gamma})=K\dot{\gamma}^{n-1}$, where $K=\left(\eta_0-\eta_{\infty}\right) \lambda_{c a}^{n-1}$.} for another comparison shown in Fig.~\ref{fig3}B The simulation results were checked with a mesh independence test.

\section{Simulation of viscoelastic flow in a 4:1 smooth contraction channel}
\label{appendix_simulation}
\subsection{Simulation framework}
Numerical simulations of viscoelastic fluid flow in a 4:1 smooth contraction channel were carried out using the viscoelastic flow module in \textsc{COMSOL} Multiphysics. The viscoelastic fluid was modeled with the modified FENE-CR constitutive equation (a finitely extensible nonlinear elastic (FENE) model) introduced by Chilcott and Rallison~\cite{chilcott1988creeping}.
The set of governing equations consists of the conservation of mass and momentum, coupled to the evolution equation of the conformation tensor:
\begin{align}
\nabla \cdot \mathbf{u} &= 0, \label{eq:continuity} \\
\rho \left( \frac{\partial \mathbf{u}}{\partial t} + \mathbf{u} \cdot \nabla \mathbf{u} \right) &= -\nabla p +  {\nabla \cdot [\eta_s\left(\nabla\mathbf{u}+(\nabla\mathbf{u})^{T}\right) ]}+ \nabla \cdot \boldsymbol{\tau}_p, \label{eq:momentum} \\
\boldsymbol{\tau}_p &= {\frac{\eta_p}{\lambda}f(\mathrm{tr}\,\mathbf{A}) \left(  \mathbf{A} - \mathbf{I} \right), }\label{eq:stress} \\
\frac{\partial \mathbf{A}}{\partial t} + \mathbf{u} \cdot \nabla \mathbf{A} - \mathbf{A} \cdot \nabla \mathbf{u} - (\nabla \mathbf{u})^{T} \cdot \mathbf{A} &= {-\frac{f(\mathrm{tr}\,\mathbf{A})}{\lambda} \left(  \mathbf{A} - \mathbf{I} \right),} \label{eq:conformation} \\
f(\mathrm{tr}\,\mathbf{A}) &= \frac{L^2 - 3}{L^2 - \mathrm{tr}\,\mathbf{A}}, \label{eq:fene}
\end{align}
where $\mathbf{u}$ is the fluid velocity, $p$ is the fluid pressure, $\rho$ is the fluid density, $\eta_s$ is the solvent viscosity, $\eta_p$ is the polymeric viscosity, $\lambda$ is the relaxation time, $\boldsymbol{\tau}_p$ is the polymer stress, $\mathbf{A}$ is the polymer conformation tensor, $\mathbf{I}$ is the identity tensor, and $L$ is the finite extensibility parameter. To account for the slight shear thinning observed in rheological measurements, the solvent viscosity $\eta_s$ is modified as a shear-rate-dependent effective value:
\begin{equation}
    \eta_s=\eta_s(\dot{\gamma}) = \eta(\dot{\gamma}) - \eta_p ,
\end{equation}
with $\eta(\dot{\gamma})$ described by a Carreau model~\cite{carreau1972rheological}:
\begin{equation}
    \eta(\dot{\gamma}) = \eta_\infty 
+ \left(\eta_0 - \eta_\infty\right)\,\left[1+\left(\lambda_{\mathrm{ca}}\dot{\gamma}\right)^2\right]^{\tfrac{n-1}{2}} ,
\end{equation}
where $\eta_0$ and $\eta_\infty$ are the zero- and infinite-shear-rate viscosities, respectively, $\lambda_{\mathrm{ca}}$ is the Carreau time constant, and $n$ the power-law index. The magnitude of the shear rate is given as
\begin{equation}
    \dot{\gamma} = \sqrt{2\,\mathbf{E}:\mathbf{E}}, \qquad 
\mathbf{E} = \frac{1}{2}\left(\nabla\mathbf{u}+(\nabla\mathbf{u})^{T}\right),
\end{equation}
where $\mathbf{E}$ is the rate-of-strain
tensor.
This construction ensures that the total apparent viscosity in the momentum equation is
\begin{equation}
    \eta_{\mathrm{app}}(\dot{\gamma}) = \eta_s(\dot{\gamma}) + \eta_p,
\end{equation}
so that the flow reproduces the experimentally observed shear-thinning behavior while the viscoelastic stresses remain governed by the FENE-CR model. All parameters $(\eta_0,\eta_\infty,\lambda_{\mathrm{ca}},n)$ are obtained from independent rheological fits, ensuring consistency between experiments and simulations, namely $\eta_0 = 2.284$~Pa s, $\eta_{\infty} = 1$~Pa s, $\lambda_{ca} = 0.4$~s, and $n = 0.919$, as shown in Fig. \ref{fig2}B.

Although experimental measurements indicate that $\eta_s$ remains nearly constant, it is in fact the polymeric contribution $\eta_p$ that should reflect shear thinning, as described in the White--Metzner model. However, this inevitably raises the question of whether the relaxation time $\lambda$ should also be treated as a function of $\dot{\gamma}$, and moreover, the White--Metzner framework cannot capture finite extensibility effects. Therefore, within the present simulation framework, we simplify the shear-thinning effect as an equivalent modification of the solvent contribution, and accordingly {take a constant polymer viscosity of $\eta_p=\eta_0-\eta_s = 0.804~\mathrm{Pa\,s}$}.

In the simulations, the value of the finite extensibility parameter $L$ was likewise obtained from experimental measurements \cite{Hu2025} and inferred from calculations. In our previous work, dimensional effects observed in CaBER experiments allowed us to determine $\lambda$ and $L^2$ through a cross-constraint approach \cite{Hu2025}. From these measurements, we estimated $L^2 \approx 2000$ for polyisobutylene (PIB) with a molecular weight $M_w = 4.8 \times 10^6$. Since $L^2$ scales proportionally with $M_w$, a reasonable estimate for the present experiments with $M_w = 2.84 \times 10^6$ PIB is $L^2 \approx 1333$, corresponding to $L \approx 36$. This value was therefore adopted in the simulations.

\begin{figure}
\centering
\includegraphics[width=0.6\textwidth]{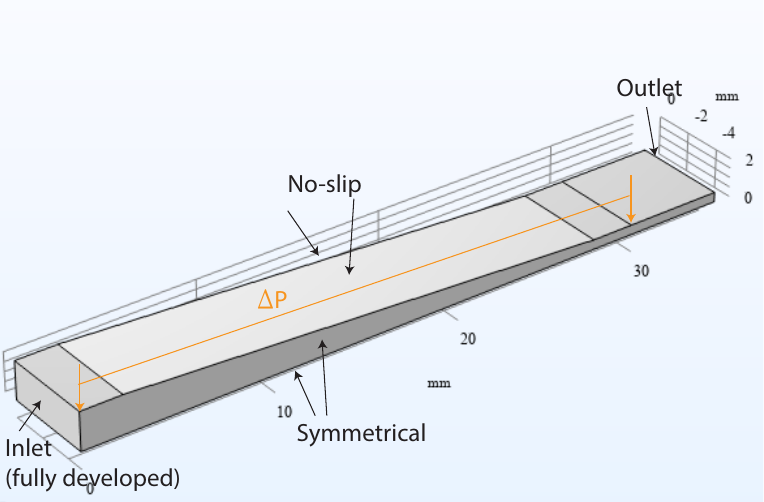}
\caption{Schematic of the finite-element simulation for the pressure drop calculation in a 4:1 smooth contraction channel. The channel dimensions are identical to those of the experimental apparatus, and the reported pressure drops correspond to the locations of the pressure sensor centers used in the experiments.}
\label{simulation}
\end{figure}

To reduce the computational cost of the three-dimensional simulations, only one-quarter of the domain was modeled by exploiting geometric symmetry, as illustrated in Fig.~\ref{simulation}. At the inlet, a fully developed velocity profile was imposed with flow rates ranging from $2.5$ to $25~\mathrm{ml/min}$, corresponding to $Q/4$ of the experimental conditions. The outlet boundary condition was set to zero static pressure, the channel walls were treated as no-slip, and symmetry planes were prescribed with zero normal derivative. Two different mesh resolutions containing 27,432 and 431,651 elements, respectively, were tested. The pressure drop difference was only 1--2\%, which demonstrated the mesh independence.

\subsection{mFENE-CR model features}
\begin{figure}
    \centering
    \includegraphics[width=0.8\linewidth]{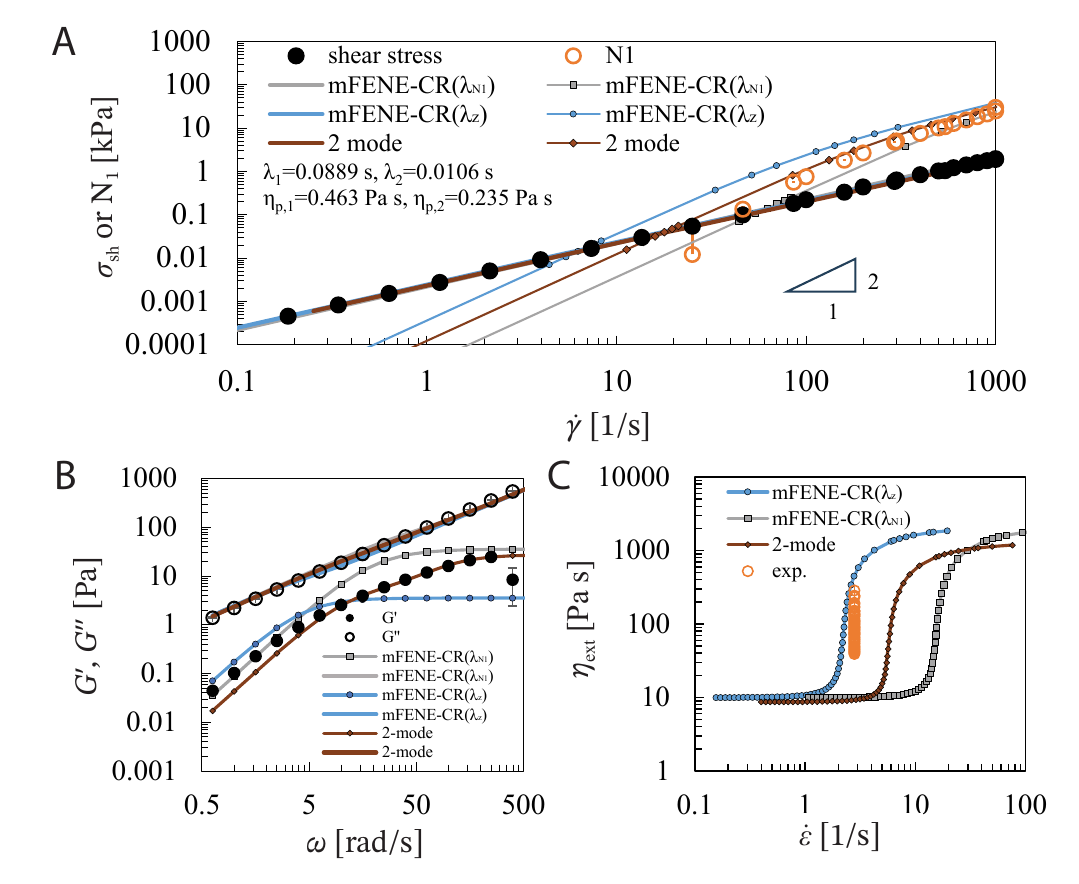}
    \caption{ Comparison between the viscoelastic model and measured rheological properties: (A) steady-shear measurements of shear stress $\sigma_{\mathrm{sh}}$ and $N_1$; (B) SAOS measurements of $G'$ and $G''$; and (C) extensional-flow measurements of $\eta_{\mathrm{ext}}$ obtained using the CaBER method, where experimental data was determined by $\eta_{\mathrm{ext}}=2\gamma/\dot{\epsilon}\mathcal{D}(t)$ and $\dot{\epsilon}=-2\mathrm{d}\ln{\mathcal{D}(t)}/\mathrm{d}t$ based on the elastocapillary regime of CaBER in Fig.~\ref{CaBER}B. $\lambda_z$ denotes the relaxation time obtained from Zimm-model fitting of the modulus data, and $\lambda_{N1}$ denotes the relaxation time calculated from $\lambda_{N1} = N_1/(2\eta_p\dot{\gamma}^2)$ based on the measured $N_1$ data, where $\eta_p=\eta_0-\eta_s$ can be determined by separate measurements of solution viscosity $\eta_0$ and solvent viscosity $\eta_s$. “2-mode” refers to the two relaxation times determined by fitting $G'$ and $G''$ with a two-mode model; higher-mode models (e.g., three-mode) fail to adequately fit the data.}
    \label{mFENE_CR_features}
\end{figure}

Although the FENE-CR model is not fully consistent with the microscopic FENE dumbbell theory, it remains widely used in practice because it provides numerical robustness and a convenient framework for capturing finite-extensibility effects while maintaining a constant shear viscosity. For these practical reasons, and given its extensive use in prior studies \cite{Szabo1997, LpezAguilar2016, TamaddonJahromi2016}, we adopt the FENE-CR model here as a representative constitutive model.

However, the observations of $\lambda_z \approx \lambda_{\mathrm{caber}} \gg \lambda_{N1}$ raise an important question: how should the relaxation time be selected for mFENE-CR model in this work, and can the use of multiple modes resolve this discrepancy? To address this, we implemented both $\lambda_z$ and $\lambda_{N1}$ in the single-mode model, and additionally determined a two-mode relaxation spectrum by fitting the measured $G'$ and $G''$ data (higher-order spectra, e.g., three modes, fail to yield meaningful fits). The corresponding model predictions are compared with experimental measurements in Fig.~\ref{mFENE_CR_features}.

Fig.~\ref{mFENE_CR_features}A shows that all three models accurately capture the shear stress $\sigma_{\text{sh}}$, and thus the shear viscosity $\eta$. However, using $\lambda_z$ leads to an overprediction of $N_1$ by approximately one order of magnitude, while the two-mode fit reduces the deviation but still overestimates $N_1$ by about a factor of five. Fig.~\ref{mFENE_CR_features}B demonstrates that a single-mode FENE model (which is effectively equivalent to the Oldroyd-B model in the small-amplitude limit) cannot reproduce the high-frequency moduli. Finally, Fig.~\ref{mFENE_CR_features}C shows that only the model employing $\lambda_z$ is able to reasonably predict the extensional viscosity.

Therefore, we acknowledge that using the Oldroyd-B or FENE models to characterize the complicated flow featuring both shear and extensional components, namely, the contraction and expansion flow in this work, inherently introduces quantitative errors. Nevertheless, this limitation does not affect the central discrepancy discussed in the manuscript, namely the opposing trends observed between pressure drop measurements and theoretical predictions in previous studies. As shown in Fig.~\ref{fig4}A, we have supplemented the simulations using both $\lambda_z$ and $\lambda_{N1}$, and found that both sets of predictions agree reasonably well with the experimental measurements, with $\lambda_{N1}$ performing slightly better. However, this leads to a substantial shift, approximately one order of magnitude, in the definition of the Deborah number, as illustrated in Fig.~\ref{fig4}B.

\section{Supplementary experimental results of pressure measurements and stability analysis }
\begin{figure}
\centering
\includegraphics[width=0.8\textwidth]{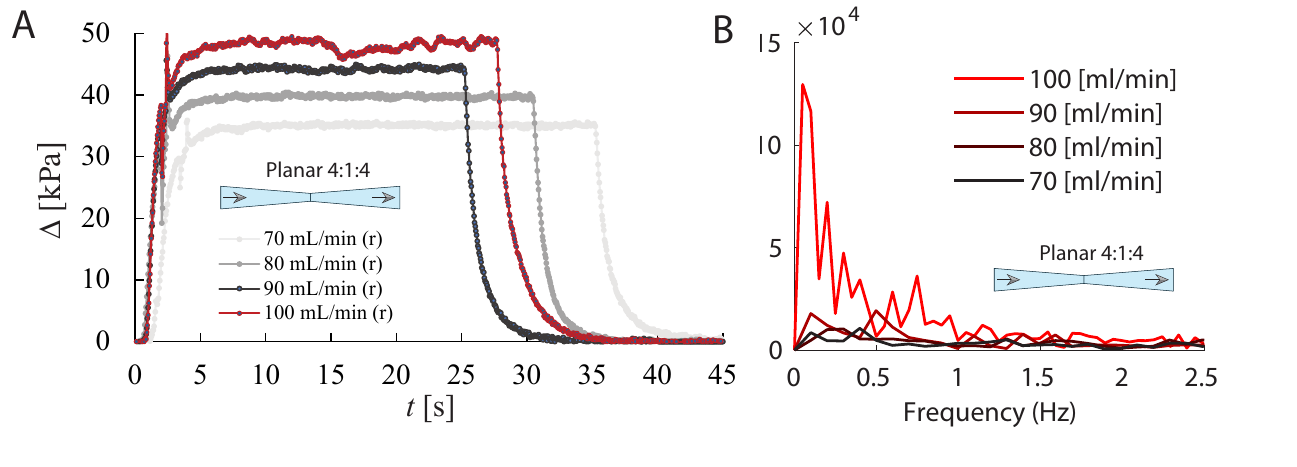}
\caption{Pressure measurements in {planar} smooth contraction--expansion ($4:1:4$) channels $\{H_{\rm en},h,H_{\rm ex}\}=\{4,1,4\}$ mm. (A) Sensor reading difference $\Delta$ and 
(B) FFT spectrum of the sensor signal, suggesting that the instability occurs at $Q=100$ ml/min.}
\label{S2}
\end{figure}

\begin{figure}
\centering
\includegraphics[width=0.8\textwidth]{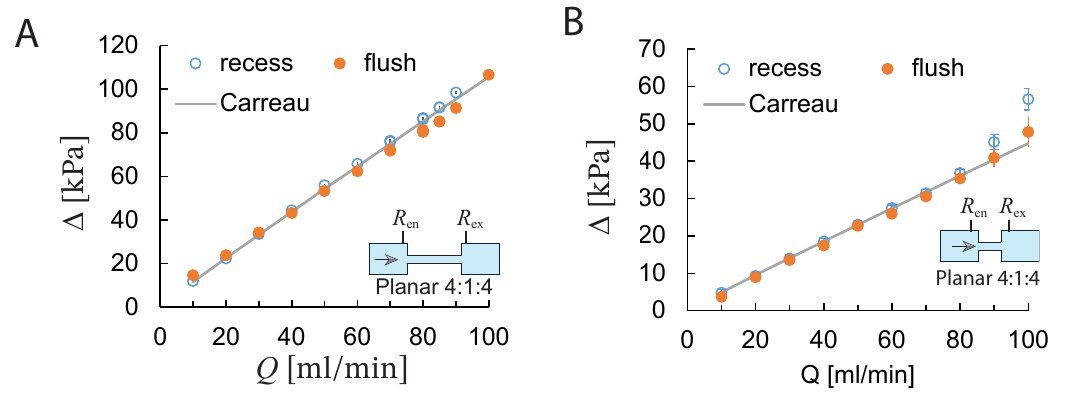}
\caption{Reading difference $\Delta$ in a {planar} $4{:}1{:}4$ contraction--expansion channel flow with different connecting length for (A) $\{H_{\rm en},h,H_{\rm ex},l_c\}=\{4,1,4,25.6\}$ mm and $\{H_{\rm en},h,H_{\rm ex},l_c\}=\{4,1,4,10.6\}$ mm.}
\label{S4}
\end{figure}

\begin{figure}
\centering
\includegraphics[width=\textwidth]{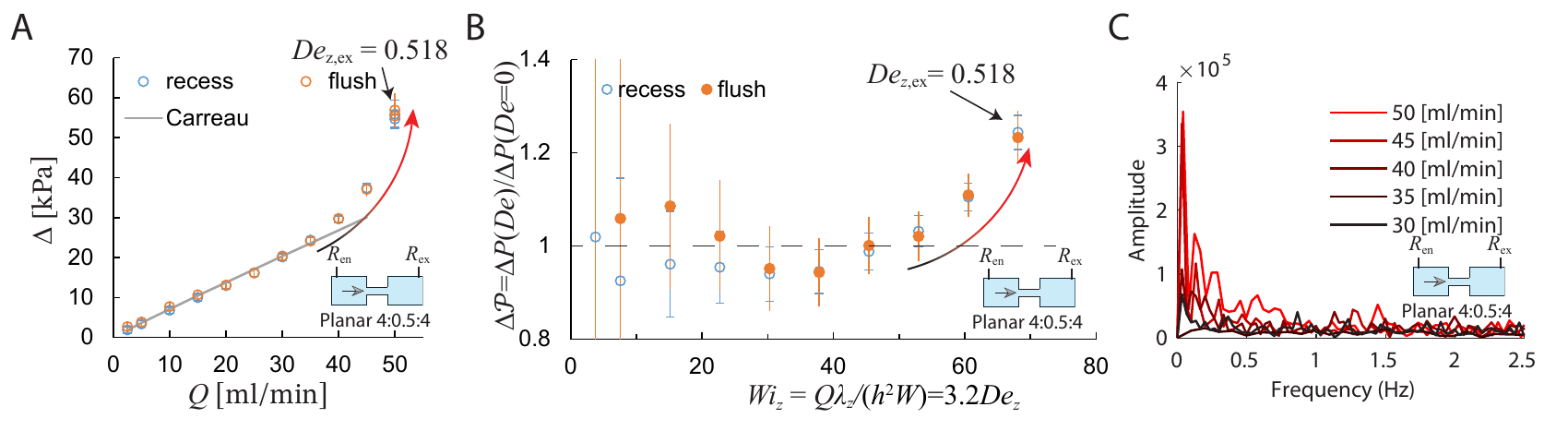}
\caption{Pressure measurements in {planar} abrupt contraction--expansion ($4:0.5:4$) channels $\{H_{\rm en},h,H_{\rm ex},l_c\}=\{4,0.5,4,1.6\}$ mm. (A) Sensor reading difference $\Delta$, (B) normalized pressure drop $\mathcal{P}$ as a function of the flow rate $Q$ or Weissenberg number ($Wi_z$), and (C) FFT spectrum of the flush sensor signal, also suggesting $\Delta\mathcal{P}>1$ associated with the instability.}
\label{S5}
\end{figure}

\begin{figure}
\centering
\includegraphics[width=\textwidth]{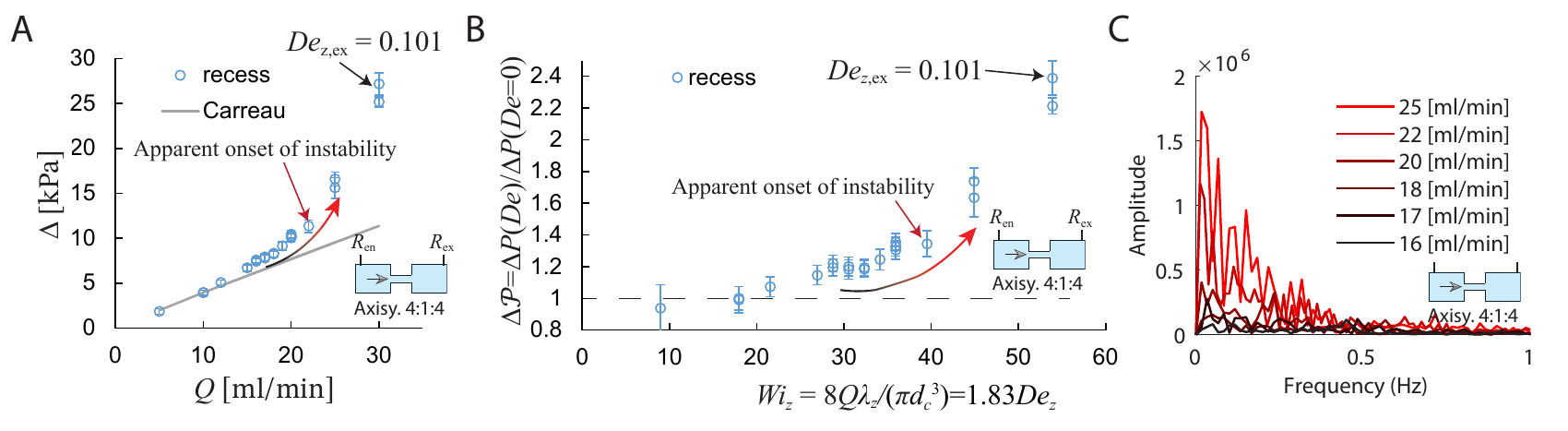}
\caption{Pressure measurements in {axisymmetric} abrupt contraction--expansion ($4{:}1{:}4$) channels with geometry $\{D_{\rm en}, d_c, D_{\rm ex}, l_c\} = \{7, 1.75, 7, 1.6\}$ mm, where $D$ and $d$ denote diameters at different positions. (A) Sensor reading difference $\Delta$ and (B) normalized pressure drop $\mathcal{P}$ as a function of flow rate $Q$ or Weissenberg number $Wi$. Since $De_{\rm z,ex} = 0.101$, the hole pressure effect can be considered canceled, and the recessed pressure sensor therefore reports the actual pressure drop.
 (C) FFT spectrum of the flush sensor signal, also suggesting an apparent instability begins at $Q=22$~ml/min, i.e., $Wi_z\approx40$. Up to this point, some data still suggest $\mathcal{P}>1$ for $20\lesssim Wi_z\lesssim40$, indicating an apparent enhancement of pressure drop even under seemingly stable conditions. This behavior is consistent with widely cited experiments \cite{Rothstein1999,Rothstein2001}, where the relaxation time used for Zimm-model fitting ($\lambda_z=3.24$~s) differed from that employed to define the Weissenberg number ($\lambda_{N1}=0.146$~s) \cite{Rothstein1999}. Using $Wi_z$ as defined in \cite{Rothstein1999}, the onset of $\mathcal{P}>1$ occurs at $Wi_z\approx 12$, with apparent instabilities emerging near $Wi_z\approx 55$.}
\label{S6}
\end{figure}

This appendix presents supplementary pressure measurements in several planar and axisymmetric contraction--expansion geometries, together with stability analyses based on FFT spectra of the pressure signals, including Fig.~\ref{S2}, Fig.~\ref{S3}, Fig.~\ref{S4}, Fig.~\ref{S5} and Fig.~\ref{S6}. These results further indicate the dependence of the measured pressure response on channel geometry, sensor configuration, and the onset of flow instabilities.

\section{Systematic measurements of an additional dilute polymer solution with varied solvent viscosity and polymer concentration}
\label{appendix_another_fluid}

To further test the robustness of our conclusions, we performed systematic measurements using an additional dilute polymer solution composed of 1.27~MDa PIB dissolved in a PB--oil solvent mixture (PB mass fraction is 38~wt.\%), where PIB and PB also denote polyisobutylene and polybutene ($M_w=2300$ Da), respectively. Compared with the solution reported in the main text, this solution differs in polymer molecular weight, polymer mass fraction, and the PB-to-oil ratio of the solvent.

\begin{figure}
\centering
\includegraphics[width=\linewidth]{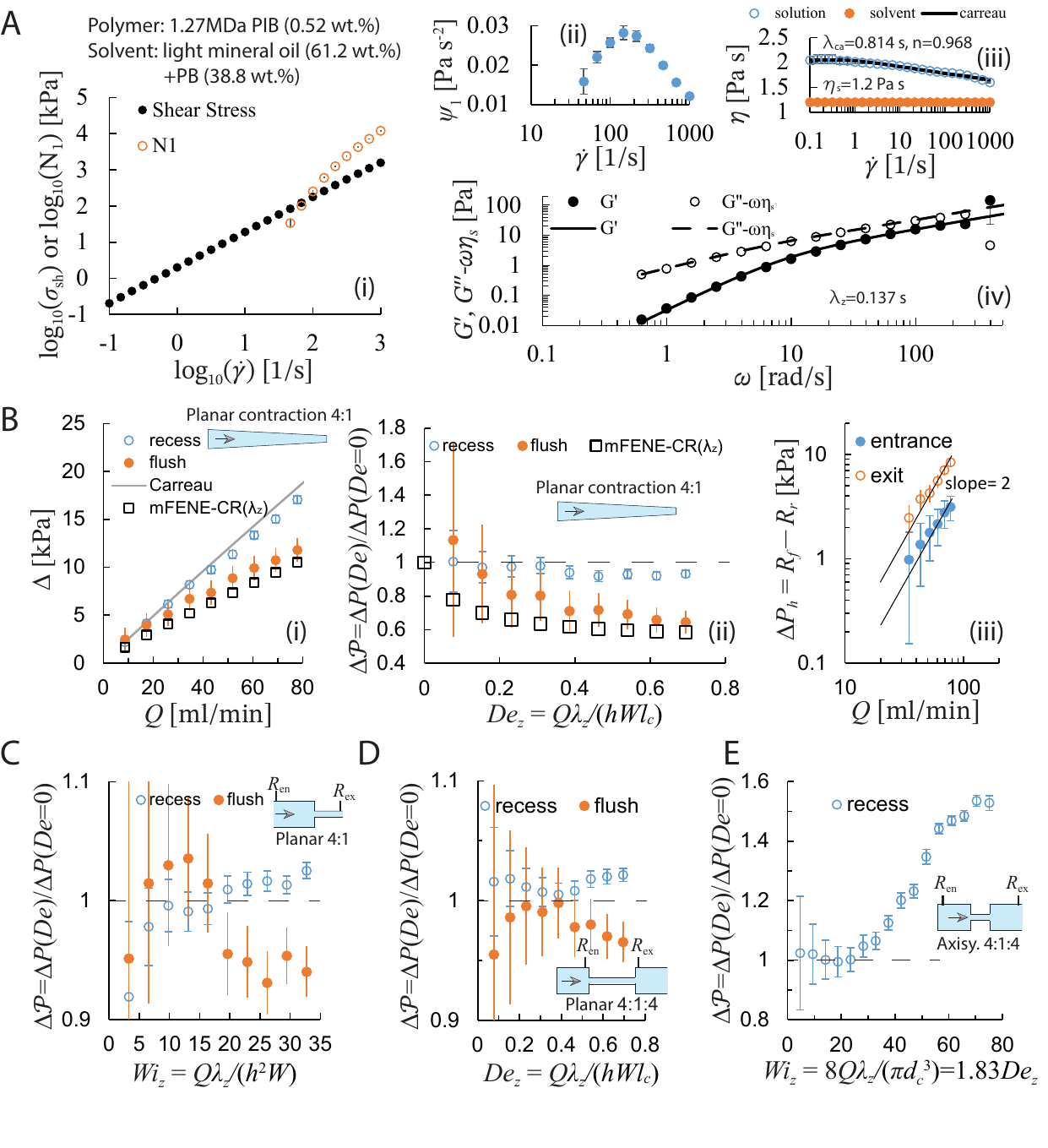}
\caption{Rheological properties and pressure-drop measurements of the 1.27~MDa PIB--PB--oil solution (here $M_w=1.27$~MDa for PIB and $M_w=2300$~Da for PB). Panel A shows the solution composition and rheological characterizations, including (i) stress--shear-rate response, (ii) first normal stress coefficient, (iii) shear viscosity, and (iv) SAOS measurements. Panels B--E show pressure-drop measurements for the smooth contraction, abrupt contraction, abrupt contraction--expansion, and axisymmetric contraction--expansion geometries, respectively.}
\label{SI_add_128PIB}
\end{figure}

The rheological characterization gives $\eta_0=2$~Pa~s, $\eta_s=1.2$~Pa~s, $\eta_p=\eta_0-\eta_s=0.8$~Pa~s, and $\lambda_z=1.37$~s. The finite extensibility is estimated as $L\approx24$ using the scaling $L\sim \sqrt{M_w}$, based on $L\approx36$ for PIB with $M_w=2.84$~MDa. As shown in Fig.~\ref{SI_add_128PIB}, the qualitative trends in smooth contraction, abrupt contraction, abrupt contraction--expansion, and axisymmetric contraction--expansion geometries are consistent with those reported in the main manuscript. In particular, the flush-mounted sensor measurements qualitatively agree with theoretical predictions, and the smooth-contraction case again shows good quantitative agreement. A slight $\Delta \mathcal{P}<1$ is also observed for the smooth contraction when using recessed sensors, which is attributed to the relatively small $\Psi_1$ of this solution.

\section{Summary of the claimed discrepancy and advances in understanding}
\label{appendix_summary_status_discrepancy}
This appendix summarizes the main claimed discrepancies between theoretical or numerical predictions and experimental pressure-drop measurements in viscoelastic contraction and contraction--expansion flows. By comparing the original discrepancy statements with related experiments and the present measurements, Table~\ref{tab:exp_summary} clarifies which issues are resolved, partially resolved, or still require further high-accuracy measurements, particularly with respect to sensor configuration, hole-pressure effects, flow stability, Reynolds number, and the relaxation time used to define $Wi$.

\begin{table*}[htbp]
\centering
\begin{threeparttable}
\caption{Summary of discrepancy (C: contraction, E: expansion).}
\label{tab:exp_summary}

\begin{tabularx}{\textwidth}{lXXXX}
\toprule
Geometry & Discrepancy statements & Related experiments & This work  & Status \\
\midrule

Planar abrupt C 
& Simulation predicts $\Delta \mathcal{P} < 1$ (Oldroyd-B, PTT), contrary to experiments\tnote{1} (Alves et al. \cite{Alves2003})
& Nigen \& Walters \cite{Nigen2002}: $\Delta \mathcal{P} \approx 1$ for planar contractions ratio 2--40

\textit{Comment:} measurement issues\tnote{a}
& Both abrupt and smooth 4:1 planar contractions show $\Delta \mathcal{P} < 1$ using flush sensors ($Re \ll 1$)
& Resolved \\

Axisym. abrupt C
& Excess pressure drop than predictions\tnote{2} (Boger~\cite{Boger1987})
& Nigen \& Walter~\cite{Nigen2002}, James et al.~\cite{James2023}: $\Delta \mathcal{P} > 1$ for axisymmetric contraction

\textit{Comment:} measurement issues\tnote{a}
& Not verified here; James et al.  \cite{James1990,James1990_2} report $\Delta \mathcal{P} < 1$ at $Re \ll 1$\tnote{b}
& Resolved by \cite{James1990,James1990_2} \\

Planar abrupt C--E
& Extra pressure drop not captured by FENE-P \cite{Keiller1993} or FENE-CR \cite{Szabo1997} simulations\tnote{3}

\textit{Comment:} $\Delta \mathcal{P} < 1$ at small $Wi$, $\Delta \mathcal{P} > 1$ at $Wi>10$ \cite{Zografos2020}
& Rodd et al.\ \cite{Rodd2005,Rodd2007}: $\Delta \mathcal{P} > 1$

\textit{Comment:} $Re > 1$ in experiments
& At $Re \ll 1$, abrupt 4:1:4 contraction shows $\Delta \mathcal{P} < 1$ at small $De$ ($Wi$)
& Resolved \\

Axisym. abrupt C--E
& Extra pressure drop not captured by FENE-P \cite{Keiller1993} or FENE-CR \cite{Szabo1997} simulations\tnote{3}

\textit{Comment:} $\Delta \mathcal{P} < 1$ at small $Wi$, $\Delta \mathcal{P} > 1$ at $Wi>10$ \cite{Szabo1997}
& Rothstein \& McKinley \cite{Rothstein1999,Rothstein2001}: $\Delta \mathcal{P} \approx 1$ first, then $\Delta \mathcal{P} > 1$ at specific $Wi$

\textit{Comment:} recessed sensors remove hole pressure effect but may mask viscoelasticity at small $Wi$; $\Delta \mathcal{P}$--$Wi$ depends on relaxation time $\lambda$
& Results consistent with far-field sensor configuration \cite{Rothstein1999,Rothstein2001}; no discrepancy for $\Delta \mathcal{P}>1$ at large $Wi$ when defining $Wi$ using $\lambda$ from CaBER or SAOS
& Partially resolved\tnote{c}; $\Delta \mathcal{P}<1$ regime needs further accurate measurements using flush sensors\\

Planar smooth C&  {Theory and simulation predict $\Delta \mathcal{P} < 1$.  Hypothesized discrepancy similar to planar and axisymmetric abrupt contractions}
& {No available experiments in planar smooth contractions before this work}
& $\Delta \mathcal{P} < 1$ using flush sensors
& Qualitatively verify theory \cite{Boyko2022,Boyko2024,Hinch2024,Mahapatra2025}\\
\bottomrule
\end{tabularx}

\begin{tablenotes}
\footnotesize
\item[1] Original statement ``Pressure recovery at high elasticity can be attributed to the lack of
dissipative effects in the Oldroyd-B model, it is contrary to experimental observations.'' 
\item[2] Original statement ``excess entry pressure drops that are significantly higher than those expected and now predicted for Newtonian or inelastic shear-thinning fluids''
\item[3] Original statement ``drop above, even in the steady regime, is not captured by simulations
that use nonlinear dumbbell models such as FENE-P or FENE-CR, with the simulations predicting a pressure drop that
decreases with increasing $Wi$ for large (and realistic) values of $L$.
\item[a] Measurements may include artifacts such as gas-driven flow or improper outlet pressure definition.
\item[b] Four flush pressure sensors are used to measure the pressure in an axisymmetric hyperbolic channel, where careful design was conducted to match the flat sensor diaphragm and the curved wall to minimize the hole pressure effect. 
\item[c] $\Delta \mathcal{P} > 1$ in the stable regime is supported by simulations (Szabo 1997; Koppol et al. 2009) at $Wi > 10$ when using the longest relaxation time. The $\Delta \mathcal{P} < 1$ regime requires further validation with high-accuracy flush sensors positioning nearby the contraction--expansion part, by which the trends of $\Delta \mathcal{P}-Wi$ can be revealed from the current $\Delta \mathcal{P}\approx1$ regime in this work and other work \cite{Rothstein1999,Rothstein2001}.
\end{tablenotes}

\end{threeparttable}
\end{table*}

\clearpage

\nocite{*}
\bibliographystyle{apsrev4-2}
\bibliography{reference}

\end{document}